\documentclass[traditabstract]{aa}  

\usepackage{graphicx}
\usepackage{tabulary}
\usepackage{txfonts}
\usepackage{lscape}
\usepackage{natbib}
\bibpunct{(}{)}{;}{a}{}{,} 
\begin{document}

   \title{Globular cluster clustering and tidal features around ultra compact dwarf galaxies in the halo of NGC\,1399 \thanks{Based on observations collected at the European Southern Observatory, (ESO Programme 076.B-0520)} }

   \author{Karina Voggel  \inst{1}
          \and \,
          Michael Hilker \inst{1}
          \and \,
          Tom Richtler \inst{2}
          }

   \institute{European Southern Observatory, Karl-Schwarzschild-Str.~2, 85748 Garching bei M\"unchen, Germany,  \email{kvoggel@eso.org}  \and Departamento de Astronom\'{\i}a, Universidad de Concepci\'on, Concepci\'on, Chile}

   \date{Received 28 July 2015; accepted 16 October 2015}

\titlerunning{Globular cluster clustering and tidal features around ultra compact dwarfs in the halo of NGC\,1399}
 \authorrunning{K.~Voggel et al.}
 
\abstract{
We present a novel approach to constrain the formation channels of
Ultra-Compact Dwarf Galaxies (UCDs). They most probably are an inhomogeneous
class of objects, composed of remnants of tidally stripped dwarf elliptical
galaxies and star clusters that occupy the high mass end of the globular
cluster luminosity function. We use three methods to unravel their nature:
1) we analysed their surface brightness profiles, 2) we carried out a direct
search for tidal features around UCDs and 3) we compared the spatial
distribution of GCs and UCDs in the halo of their host galaxy. 

Based on FORS2 observations under excellent seeing conditions, we have studied
the detailed structural composition of a large sample of 97 UCDs in the halo
of NGC\,1399, the central galaxy of the Fornax cluster, by analysing their
surface brightness profiles. We found that 13 of the UCDs were resolved above
the resolution limit of 23\,pc and we derived their structural parameters
fitting a single S\'ersic function. When decomposing their profiles into
composite King and S\'ersic profiles, we find evidence for faint stellar
envelopes at $\mu=\sim 26 \rm mag\,arcsec^{-2}$ surrounding the UCDs up to an
extension of 90\,pc in radius. 

We also show new evidence for faint asymmetric structures and tidal tail-like
features surrounding several of these UCDs, a possible tracer of their origin
and assembly history within their host galaxy halos. In particular, we present
evidence for the first discovery of a significant tidal tail with an extension
of $\sim$350\,pc around UCD-FORS\,2.

Finally, we studied the local overdensities in the spatial distribution of
globular clusters within the halo of NGC\,1399 out to 110\,kpc, to see if they
are related to the positions of the UCDs. We found a local overabundance of
globular clusters on a scale of $\leq$1\,kpc around UCDs, when we compare it to the
distribution of globulars from the host galaxy. This effect is strongest for
the metal-poor blue GCs. We discuss how likely it is that these clustered
globulars were originally associated with the UCD, either as globular 
cluster systems of a nucleated dwarf galaxy that was stripped down to its nucleus,
or as a surviving member of a merged super star cluster complex.
} 

   \maketitle
%

\section{Introduction}

Ultra-Compact Dwarf Galaxies (UCDs) were first discovered in the Fornax cluster (\citealt{Minniti1998, Hilker1999, Drinkwater2000}). Although their name implies a galaxy origin, their nature and origin is still puzzling more than 15 years after their discovery. Their sizes (3-100\,pc) and luminosities $(-10<M_{\rm V}<-14)$ have filled the void in the scaling relations of early-type stellar systems (\citealt{Misgeld2011, Norris2014}). The size and magnitude gap in between classical globular clusters (GCs) and dwarf galaxies was not populated before the advent of UCDs. 
 
Two main scenarios for the origin of UCDs are being discussed in the literature: 1) UCDs are the surviving nuclei of tidally stripped (dwarf) galaxies (\citealt{Bekki2003}, \citealt{Pfeffer2013}) and 2) UCDs are the bright end extension of the globular cluster luminosity function, either formed as genuine GCs (\citealt{Murray2009, Mieske2004}) or are the result of a star cluster complex, where many clusters merge into a massive UCD-like object (\citealt{Fellhauer2002, Bruens2009, Bruens2011}).

The view that UCDs are an inhomogeneous class of objects with different origins, i.e. made up of high mass GCs as well as stripped nuclei, has been emphasized ever since their discovery (\citealt{Hilker2006, DaRocha2011, Brodie2011, Norris2011}). The contribution of each formation channel is not well constrained yet and is highly under debate. Identifying the origin of individual UCDs is observationally very difficult since most observed parameters do not differentiate between both formations channels. Smoking gun properties, like a massive black hole, an extended star formation history or massive tidal tails, are expensive to obtain for such barely resolved objects. One of the most convincing proofs for a stripped nucleus-origin of an individual UCD is the detection of a supermassive black hole (SMBH) in M60-UCD1 in the Virgo cluster by \cite{Seth2014}. Another very recent example is the detection of an extended star formation history in NGC\,4546-UCD1 by \cite{Norris2015}. But, for large ensembles of UCDs there is no estimator yet, which can distinguish between both UCD formation channels and thus constrain their fractions among a UCD population.

The Fornax cluster, in particular its central galaxy NGC\,1399, has a well studied UCD and globular cluster population (\citealt{Dirsch2003, Bassino2003, Dirsch2004, Bassino2006, Schuberth2010}). NGC\,1399 hosts $\sim$6500 globular clusters within 83\,kpc (at a distance of 19\,Mpc) with a specific frequency of $\rm S_{N}=5.5$, making it a good testing ground for GC/UCD populations.

At the distances of the Virgo and Fornax clusters most UCDs are barely resolved on ground-based images under regular seeing conditions. They can however be spatially resolved by space telescopes like HST or adaptive optics supported ground based imaging. The light profiles of a few UCDs have been studied in detail via HST photometry (\citealt{Evst2007, Evst2008, dePopris2005}). It has been found that a simple single King profile does not fit the luminosity distribution of the most luminous and extended UCDs well. A King core and a Sérsic or exponential halo component are needed to explain their surface brightness profiles (\citealt{Evst2008, Strader2013}). Especially the brightest UCDs exhibit compact and concentrated profiles in their centres and  extended exponential wings with no signs of a sharp tidal truncation. Some of them exhibit envelope components with effective radii up to 100\,pc (e.g. \citealt{Drinkwater2003, Richtler2005}).
There are 6 UCDs listed in the literature, which have double surface brightness profiles: VUCD\,7 in the Virgo Cluster and UCD\,3 and UCD\,5 in Fornax (\citealt{Evst2008}); the compact object M59cO (\citealt{Chili2008}); and the very compact, SMBH-harbouring M60-UCD1 (\citealt{Strader2013}).

One of the scenarios suggested for the formation of nuclear star clusters (NSCs) is that massive globular clusters migrate towards the center of their host galaxy via dynamical friction and merge there in less than a Hubble time (\citealt{Tremaine1975, Capuzzo1993}). This process is especially efficient in dwarf galaxies. \cite{Lotz2001} found a deficit of bright globular clusters in the central regions of nucleated dEs as compared to the outer regions, which can be the result of inwards migration of massive GCs. Recently, \cite{Arca-Sedda2014} have derived, in a statistical approach, the number of surviving clusters around a galaxy as function of its mass, after a full Hubble time of dynamical friction at work. Their models predict that for a host galaxy with a stellar mass of $M=10^{10}M_{\odot}$, on average, $65\%$ of the original globular cluster population has migrated to the centre within one Hubble time. This fraction rises for smaller host masses. At $M=10^{9}M_{\odot}$ we already expect $80\%$ of the original globular cluster population to have merged into the nucleus via dynamical friction. Assuming that UCDs are the stripped nuclei of progenitor dwarf galaxies in the mass range $M=10^{9}-10^{10}M_{\odot}$, we can test whether observed UCDs show signatures of this process. In particular, we can test if there is a statistical overabundance of globular clusters in close proximity of UCDs, as we expect that inwards migration of GCs is still ongoing within the shrinking tidal radius of the disrupting dwarf galaxy. And, as the most massive globular clusters have the shortest dynamical timescales, we expect only lower mass GCs to survive around UCDs. This can be tested by sampling the luminosity function of the companion objects and see if it agrees with a GC population that is skewed towards low masses in the globular cluster luminosity function (GCLF). Also we would expect that small, low surface brightness envelopes from the progenitor galaxies are left within the present-day tidal radius of the stripped nuclei.

Merging super star cluster mergers are also expected to show substructure and non-merged companion GCs. In \cite{Bruens2011} it was shown that such a merging super star cluster complex has several surrounding close GC companions at 70\,Myrs after the start of the merging process. After 280\,Myrs of merging it is left with one GC companion. Finally after 1.3\,Gyrs no more companions or substructure are visible in that particular simulation and the merging process has finished. In another simulation of merging star cluster complexes that leads to the formation of an extended star cluster of $10^{5}M_{\odot}$ ('faint fuzzies'), \cite{Bruens2009} showed that substructure in form of non-merged star clusters can survive up to a merger age of 5\,Gyr. Also tidal tails are formed while the merged star cluster complex orbits in the gravitational field of its host galaxy. Thus we might expect to find companions around merged super star clusters up to several Gyrs after their formation.

In this work we present novel approaches to constrain the origin of UCDs in the Fornax Cluster.

Firstly, we carry out a detailed structural decomposition of the surface brightness profiles of several UCDs around NGC\,1399 using ground based, very good seeing images (sections 2 and 3). We examine their surface profiles by fitting single component models to them, as well as decomposing some of them into an envelope and core component. We compare how modelling UCDs with different light profiles affects the scaling relations of UCDs within the larger picture of early-type systems.

Secondly, we examine a large sample of confirmed UCDs for direct signatures of ongoing tidal stripping (section 4). We look for direct ``smoking gun evidence'' for the tidal transformation of UCDs by searching for tidal tails.

Finally, we search for signatures of associated companion star clusters in the distribution of GCs around UCDs, which we expect if the UCD was a nucleated dwarf galaxy which previously had its own globular cluster system or it originated from a super star cluster complex that still has substructure (section 5). We investigate how likely it is, that these clustered globulars were either originally part of the ancestor dE galaxy before it was stripped down to its nucleus, or originated from a super star cluster complex that has not fully merged yet. The main difference of the two scenarios being the age of the UCDs with substructure, as we do not expect a star cluster complex older than 5\,Gyrs to retain substructure, but rather be fully merged into one smooth object.

\section{Imaging}
\begin{figure}
   \centering
   \includegraphics[width=\hsize]{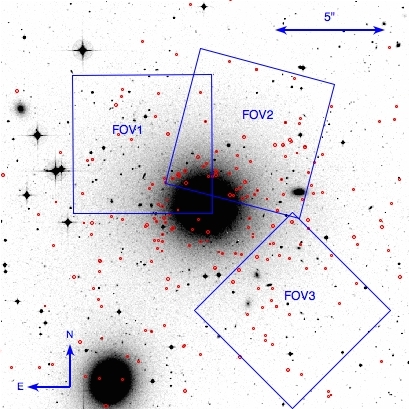}
      \caption{A DSS cutout image of $20'x20'$ around NGC\,1399 (RA=03h38m29.0s DEC=-35d27m02s), the central galaxy of the Fornax cluster. The locations of the three FORS2 pointings with $6.8'\times6.8'$ size are shown as blue squares. The confirmed Fornax UCDs/GCs within this region are marked with red circles.}
         \label{fig:FOV}
   \end{figure}

To study UCDs within the Fornax cluster, we used data from ESO programme 076.B-0520 (PI:Richtler). The imaging data were taken in the nights October 9th and 10th, 2005, with the high resolution collimator mode of FORS2, which is mounted on UT1 of the Very Large Telescope (VLT). Three separate fields were observed in the R-band. Fields 1 and 2 with 3x800\,s exposure time each and field 3 with 5x800s. Every single exposure was reduced separately with our own IDL routine. The reduction process included all standard procedures for image reduction. First, a masterbias was created from 5 bias exposures, then a normalized master flatfield of 5 separate bias-subtracted flatfields was created. The original science exposures were then corrected for bias and flatfield effects and subsequently stacked together using the IDL routine MEDARR, creating a median stacked image of all three exposures. Before stacking we fitted gaussian models to 4 bright stars in each exposure to determine their positions with sub-pixel accuracy. This was used to correct for small sub-pixel shifts in the astrometry before stacking the exposures, so that our psf size was not increased. The pixel scale of the observations is $0.126$ arcsec/pixel, which corresponds to a physical scale of 92\,pc\,arcsec$^{-1}$ at the distance of NGC\,1399, which we will assume throughout the paper as $(m-M)=31.39$ (\citealt{Freedman2001}), or 18.97\,Mpc. For each final stacked image we retrieved the psf using the GETPSF idl routine taken from the NASA IDL Astronomy Users Library (\citealt{Landsmann1993}) \footnote{http://idlastro.gsfc.nasa.gov/}. Mean FWHM-values of the PSFs in the three fields are FWHM$=$0.53\arcsec, 0.55\arcsec, 0.53\arcsec for pointings Nr.1, 2 and 3, respectively. The positions of the three pointings are shown in figure \ref{fig:FOV}. The orientations of the fields were optimized to cover most objects presented in \cite{Richtler2005}. 

\section{UCD Analysis}
\subsection{Surface brightness analysis}
\label{sec:surf}
We compiled a sample of 313 UCDs and GCs in the Fornax cluster with stellar masses above $M=10^{6}M_{\odot}$ (\citealt{Firth2007, Mieske2002, Mieske2004, Dirsch2004, Richtler2008, Schuberth2010}), which all are confirmed Fornax members according to their radial velocities. To determine their stellar masses consistently for the full sample, we used their $V$-band magnitudes and $(V-I)$ colours and calculated their $V$-band mass-to-light ratios from the simple stellar population models (SSP) of \cite{Maraston2005}. For the UCDs we assume an age of 13 Gyr and a \cite{Kroupa2001} initial mass function (IMF) with a red horizontal branch. A fit to the tabulated $M/L_V$ and $(V-I)$, valid for the colour range $0.80 < (V - I) < 1.40$, can be expressed as: 

\begin{equation}
\frac{M}{L_{V}} = 4.408 + 1.782 \cdot \rm arctan[11.367 \cdot ((V - I) - 1.162)]
\end{equation}
The root mean square of the fit is rms$=$0.101.
In figure \ref{fig:FOV} a 20'x20' cutout DSS image is shown, centered on NGC\,1399. The UCDs of the Fornax sample are marked with red circles. The three FOVs of our FORS2 observations are shown as blue squares. 97 UCDs/GCs are located in those fields.

We analyzed the surface brightness profiles of these 97 UCDs using the two dimensional fitting algorithm GALFIT (\citealt{Peng2002}). GALFIT is a parametric approach to light profile fitting, which minimizes the residuals between the two dimensional model and the image. The difference between model and original image is given as a  residual map. As the diameters of the UCDs are very close to the spatial resolution of our images it is crucial that we use a fitting technique that takes the smearing of the light profile by the PSF into account. GALFIT  convolves each analytic profile with an input point-spread function (PSF). We derived the input PSFs using the GETPSF IDL routine for, on average, 20 bright stars in each observing field and each chip separately, as the PSF varies in each observation and also per chip. The average FWHM of our derived PSFs is 4.25 pixels which corresponds to 0.55\arcsec. For each UCD a cutout of 120$\times$120 pixels, which corresponds to 15\arcsec$\times$15\arcsec, was created from the large image as an input for GALFIT. We used GALFIT to fit 2 single and 2 double luminosity profile models to each UCD: a single Sérsic (S), a single King (K), a double King+Sérsic (KS) and a Sérsic+Sérsic (SS) model. 

The Sérsic surface brightness function used to model galaxy luminosity profiles can be given as:
\begin{equation}
I(R) = I(\rm eff) \,\, exp \left[-b \left(\left(\frac{R}{R_{\rm eff}}\right)^{\frac{1}{n}} - 1 \right) \right].
\end{equation}
$R_{\rm eff}$ is the half light radius, $I(\rm eff)$ is the surface brightness at the half light radius, n is the shape parameter of the function and b is a constant which depends on n with a good approximation as  
\begin{equation}
b(n) = 2n-0.324
\end{equation}
See \cite{Ciotti1991} for exact values.

For n=1 the profile is an exponential law, and for n=4 it is a de Vaucouleurs profile.

The generalized King profile (in original form from \citealt{King1962}) is given as: 
\begin{equation}
I(R) = I_0 \left[\frac{1}{(1+(R/R_c)^2)^{\frac{1}{\alpha}}} - \frac{1}{(1+(R_t/R_c)^2)^{\frac{1}{\alpha}}} \right]^{\alpha},
\end{equation}
where $R_{\rm c}$ is the core radius and $R_{t}$ is the tidal radius where the profile is truncated. $I_{0}$ is the central surface brightness. $\alpha$ is the shape parameter of the King profile. The classical profile fixes $\alpha=2$, but we use it in its generalized form where $\alpha$ can vary. Often the concentration index c is used to characterize the profile, which is defined as $c=log(R_{\rm t}/R_{c})$. The King profile is often a good fit to light profiles of globular clusters with their flat cores and truncated outer parts.

All model parameters of our fits were allowed to vary within GALFIT. In cases where a very close faint point source or an asymmetric tidal feature was visible in the image, we fixed the ellipticity to 1.0, providing a spherical model. When there is a nearby faint substructure GALFIT often tends to settle into a local $\chi^{2}_{\nu}$ minimum with a very elliptical model, which results in a bad representation of the real object. For objects without faint substructure we allowed the ellipticity to vary. The ellipticities of the UCDs is given by GALFIT as the ratio between semi-minor and semi-major axis of the fitted model. For all UCDs where there were no faint tidal features nearby we allowed the ellipticity parameter to vary. The ellipticities from the Sérsic fits were all larger than 0.75, which shows that UCDs are quite round objects with few deviations from their round shape in the inner light profile.

As many UCDs are located close to the very bright galaxy NGC\,1399, the background level in the cutout images is not uniform but has a brightness level which depends on the distance to the main galaxy. For each UCD fit we determined the individual sky background and allowed it to have an additional sky gradient in x and y direction.

We only accept GALFIT parameters if the best fit converged with no runaway parameters, i.e. parameters whose values do not diverge with increasing radius, and if $r_{\rm eff}$ of the fitted profile is $>2 \rm pix=23.18$\,pc, which corresponds to 45\% of the size of the FWHM and 0.252\arcsec in angular size. This is the resolution limit we adapted for all our fits. For the double component fits, the $r_{\rm eff}$ values of both components are also required to lie above this resolution limit.

   \begin{figure}
   \centering
   \includegraphics[width=\hsize]{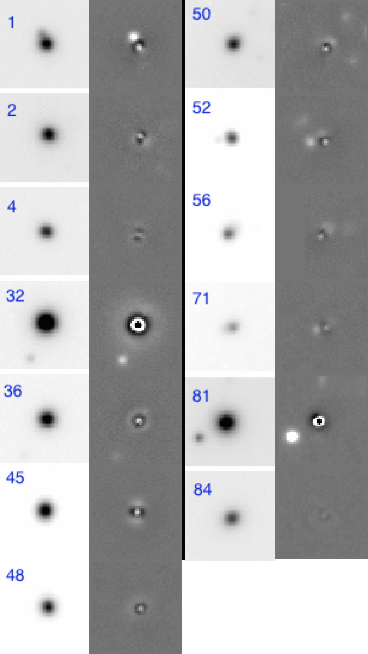}
      \caption{The left columns show the thumbnail images of the 13 UCDs that were found to have a half-light radius above 23\,pc, which we adopted as limit for a reliable measurement. The right hand column next to each UCD shows the residuals after subtracting the best fit single Sérsic model from the observed image. Each UCD is marked with its identifier number among the 97 UCDs in both FORS fields. Their properties are listed in detail in table \ref{ucddat}. }
         \label{fig:extended}
   \end{figure}  
Applying these cutoffs we find that 54 UCDs have a converged S\'ersic fit, and of these 13 have a half-light radius above the resolution limit and no runaway parameters. The cutout images of these 13 extended UCDs and their residuals of the S\'ersic fit are shown in figure \ref{fig:extended}. The structural parameters we derived from the S\'ersic fits to these UCDs are given in table \ref{ucddat}.

The quality of the fits is shown as residuals in figure \ref{fig:extended}. These residual maps are the results from subtracting the best-fit Sérsic model from each observed object. As can be seen some larger residuals are left in the central parts for most UCDs. These are either due to an undersampled PSF determination in the centres or due to and imperfect description of the Sérsic profile to the very central parts of the UCDs. For several objects we see faint outer residuals (e.g. UCD-FORS 4, 32, 36, 45). This suggests the existence of a faint underlying second component, which we will test with double profile fits. For UCD-FORS\, 81 ($=$Fornax UCD2) we find an effective radius of $r_{\rm eff}=37.82 \pm 2.58$\,pc and a Sérsic index of $n=2.03$. For this object a much higher Sérsic index of $n=6.8$ and an effective radius of 26.6\,pc was derived in \cite{Evst2007}, based on higher resolution HST photometry. High Sérsic indices are very sensitive to the determination of the sky level, and already a slight underestimation will lead to a higher Sérsic index, which might cause this difference. Another explanation for this discrepancy is that an intrinsic double profile was fitted with a single component in their work.

The average Sérsic index of our 13 extended UCDs is $n=3.82$. The objects with larger Sérsic indices are more centrally concentrated and have extended wings towards faint levels compared to those with lower Sérsic $n$ values. As mentioned above, profile fits to extended wings can be sensitive to background determinations. Therefore, it is important to determine our background carefully, including the modelling of the gradient introduced by the central galaxy NGC\,1399 to minimize these effects. 

For the King profile fits, we have 47 UCDs, for which the fit returned results, but only three of them have no runaway parameters and effective radii above the resolution limit we adapted. These three objects above the resolution limit
 showed an extremely low alpha index of  $\alpha<0.2$ which is unphysical. Thus we consider none of the UCDs to be well fitted by a single generalized King law in contrast to previous HST results. None of the UCDs have either a flat cored light profile, nor a marked truncation radius, which we would both expect from typical King profiles.

We also attempted to fit the UCDs with double component profiles, King-Sérsic (K+S) and Sérsic-Sérsic (S+S). Many of these fits did not converge and had runaway parameters. 
And none of those that converged had effective radii for the inner component that were above our resolution limit when we allowed all parameters to vary. As we know from the 5 known double profile UCDs in the literature (\citealt{Evst2007,Chili2008,Strader2013}), the inner component usually has $r_{\rm eff}<15\, \rm pc$. Finding none that has an inner component larger than 23\,pc was thus expected, within our seeing limited ground based dataset.

To test how it would affect the outer Sérsic component if we actually have an inner King component, we reran GALFIT on the 13 extended UCDs with a fixed size King component in the center. We held the King effective radius at $r=11.6$ pc with a concentration of $c=30$ and allowed the magnitude to vary. Also the Sérsic envelope profile was allowed to vary in its parameters. All these fits converged except for the one of UCD-FORS\,32 (UCD6). Instead for this UCD a double Sérsic profile is the best fit to the data. Notably this is the object with the lowest Sérsic index of $n=0.94$ and thus the most cored profile consistent with the profile shape derived by \cite{Evst2008}. 

In table \ref{KSdat} the results of the K+S fits are shown, including the relative change in $\chi^{2}$ between the Sérsic only and the K+S fit. For 11 out of the 13 UCDs the relative change of $\chi^{2}$ is positive and thus the quality of the fits increased when using a double K+S profile. For UCD-FORS\,4 and 52 these percentages are slightly negative, but with only a tenth of a percent the quality of the fit is almost unchanged. The average increase in the quality of the fits is $\Delta \chi^{2}$=10.59\%. The double profile fit is especially good for UCD-FORS\,32, 45 and 81, for which a faint envelope is even visible from the images in figure \ref{fig:extended} by eye. For illustrating the improvement in residuals for the double component fits, those three objects have been included in the last three panels of figure \ref{fig:surfprofiles} and all of them show significant improvement of the residuals when using a K+S profile instead of a single Sérsic.
The asymmetric faint features around UCD-FORS\,1, 52 and 71 are caused by faint background point sources which will be further discussed in section \ref{sec:companion}. 

The effective radii for the envelope S\'ersic components are shown in figure \ref{fig:sizemag} as dark blue circles with crosses. They are connected to the respective parameters derived by the single Sérsic fit with a solid line. 

It is obvious from the plot that the effective radii of these 'envelope' components are larger than their single component counterparts. In the scaling relations some of these envelopes are located in the empty area between the branch of the dwarf elliptical galaxies and the star cluster like objects (UCDs and GCs).  

As expected, the Sérsic index for the envelope decreased due to its less concentrated profile. As many as six of our envelopes now have Sérsic indices of 2 or smaller, which is closer to the $n=1$ exponential profile that is usually measured for dEs.

\subsection{Surface brightness profiles}
\label{sec:profiles}

\begin{figure*}
   \centering
   \includegraphics[width=18.5cm]{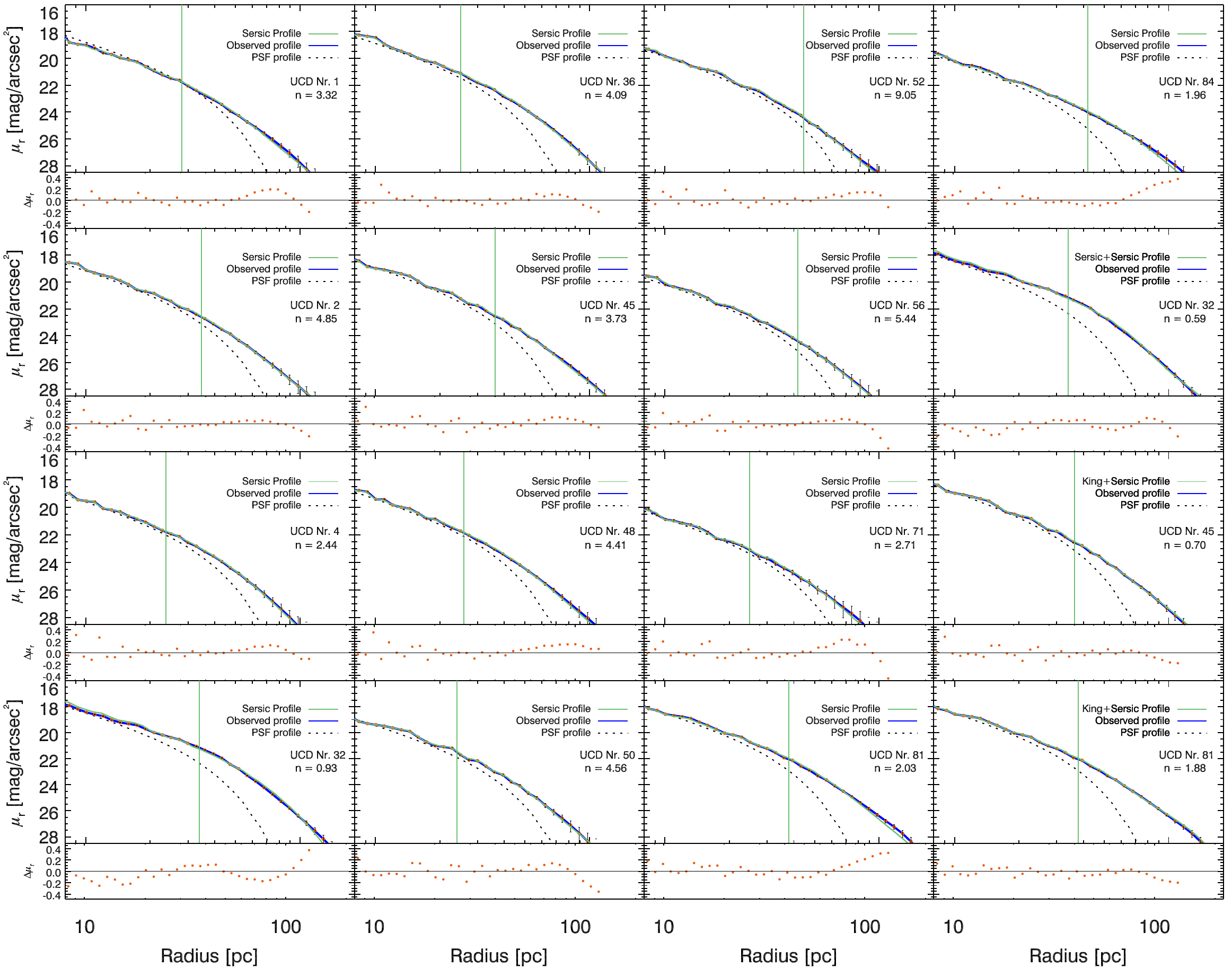}
      \caption{Here the isophotal analysis of the R-band images of 13 extended UCDs is shown. For each UCD, the top panel in each plot shows the output of the GALFIT Sérsic model (green) compared to the observed background subtracted surface brightness profile (blue line, orange points). The shape of the PSF is also plotted as dashed black line. The Sérsic index $n$ and the identification number of the UCD is noted within each plot. The lower panel of each plot displays the residuals between the model profile after convolution with the PSF and the observed profile. The vertical green line is the effective radius of the best fit Sérsic profile. The last three panels in the fourth column show the model results for the King+Sérsic fits for UCD FORS 45 and 81 respectively. For UCD-FORS \,32 the Sérsic+Sérsic result is shown. The two-component models of these three UCDs show improved residuals compared with their single profile fits.}
         \label{fig:surfprofiles}
\end{figure*}

For those 13 UCDs where we were able to measure the structural parameters directly we show their one dimensional, background corrected radial light distributions in figure \ref{fig:surfprofiles}. The last three panel of the 4th column show the best fit Sérsic+Sérsic fit for UCD=FORS\,32 and the respective King+Sérsic fits for UCD-FORS\,45 and 81. In green the best fit convolved Sérsic profile is plotted. In blue the background corrected observed surface brightness profile is shown. For this we used the IRAF ellipse task and analysed the surface brightness level of the UCDs as a function of aperture radius. This non-parametric curve-of-growth analysis does not correct for the smearing of the light profile in the central parts by the PSF but shows the profile as observed. The errorbars for the blue profile denote the change in the profile when we vary the determined background value by 5\%. The black dashed line is illustrating the surface brightness profile of the PSF. The vertical green line shows the size of the effective radius $R_{\rm eff}$ of the best fit Sérsic profile. Below each isophote plot the residuals between the model profile after convolution with the PSF and the observed profile is shown.

For the inner parts of most UCD isophote profiles (r$<50$pc) the residuals are small and scattered around zero. For several objects at larger radius the residuals become positive meaning an excess of the observed light compared to the model. For UCD-FORS\,1, 52 and 71 this can be accounted for by the very close second point sources we detected after subtracting the UCD profile (see figure \ref{fig:extended}). As also visible from the two dimensional residuals plotted in figure \ref{fig:extended}, for several UCDs (e.g. 1, 71, 84) the two- and one-dimensional residuals both show a faint excess of light at radii $r>70$pc, i.e. the Sérsic profile dropped too fast in surface brightness compare to the true light profile of the UCDs. A very interesting case is UCD-FORS\,32 whose best fit light profile has the smallest Sérsic index of the whole sample of n=0.93. The behaviour of its residuals shows an excess of observed light between 30 and 50\,pc compared to the profile, and between 70 and 100\,pc the Sérsic profile has a higher surface brightness than the object. Profiles with Sérsic indices below 1 indicate a light distribution with a central flat (core) and a truncation at larger radii. In the second panel of the 4th colum this object is shown with it's best fit Sérsic+Sérsic fit. The ``envelope" component of this fit has now an even lower Sérsic index of 0.59 and a half light radius of 88.69\,pc,  the improvement in the quality of the fit is easily visible from the residuals. The double profile results from UCD-FORS\,45 and 81 are plotted in the last two panels of the last column, and also show significant reduction in the amount of residual compared to their single-component counterpart. 

All light profiles of the 13 extended UCDs lie significantly above the PSF profile (dashed black) and are clearly not compatible with a point source profile within their errorbars.
 
In table \ref{ucddat} we summarise the structural parameters from the best fit Sérsic models for the 13 UCDs that have half-light radii above the resolution limit of 23\,pc. The $(V-I)$ colour and the radial velocities are taken from their original sources described in detail in section \ref{sec:surf}. The provided errors on the fit parameters, both in tables \ref{ucddat} and \ref{KSdat}, are those given by GALFIT and should be taken with caution, as these errors might be underestimated and do not take into account any systematic effects. For objects this close to the resolution limit we expect the true total errors to be at least 10-20\%.

\subsection{Colour magnitude diagram}
   \begin{figure}
   \centering
   \includegraphics[width=\hsize]{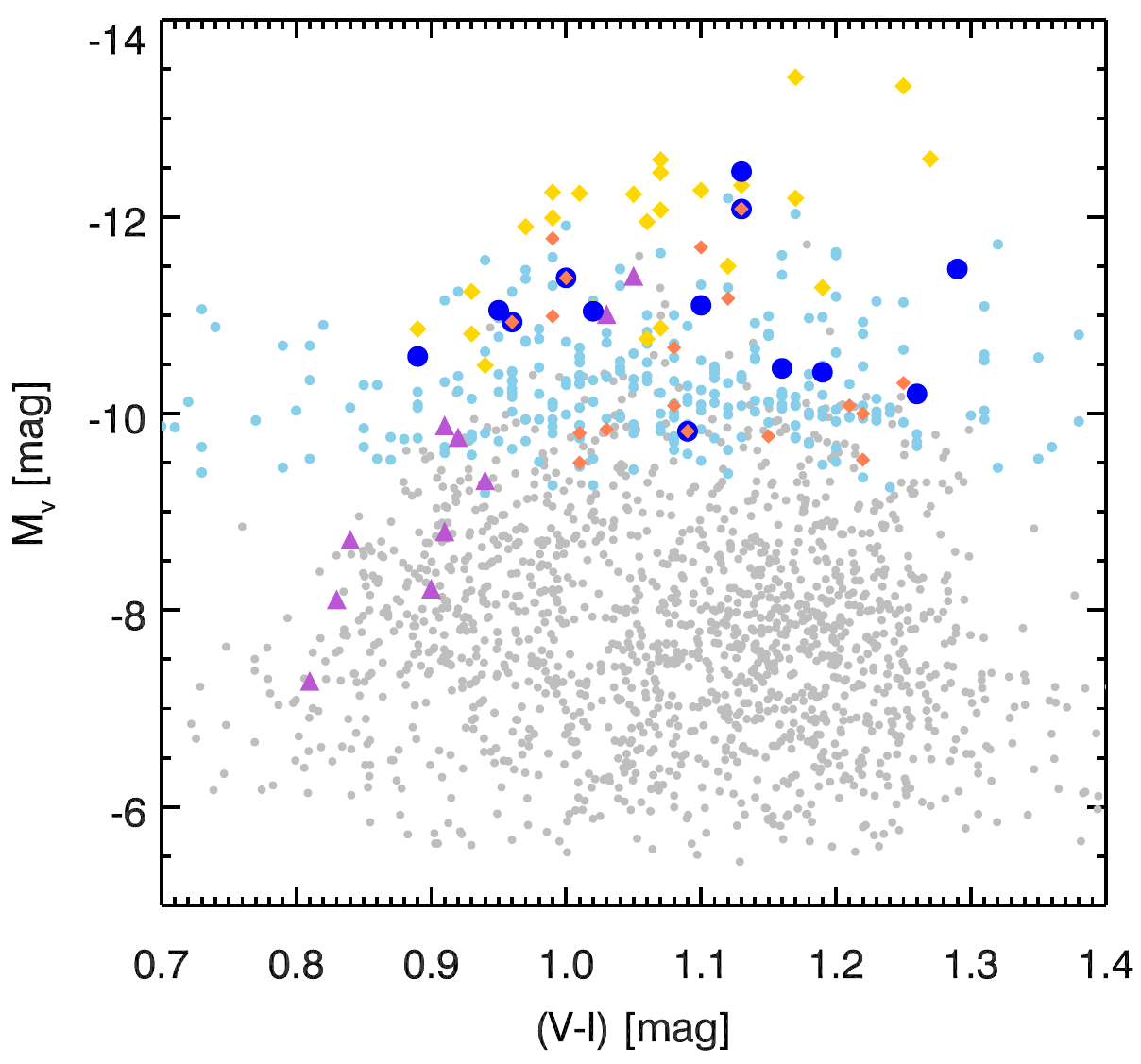}
      \caption{Colour magnitude diagram for the extended UCDs (dark blue) and UCDs with companions as orange diamond symbols. These are compared to the parameters of the rest of our NGC\,1399 UCD sample with $M> 10^{6}M_{\odot}$ in light blue. The yellow diamonds mark UCDs from the work of \cite{Evst2008}. The light grey dots are the NGC\,1399, NGC\,1404 and NGC\,1389 GCs from the Fornax ACS Survey (\citealt{Jordan2007}). The purple triangles are nuclei from dE galaxies in the Fornax cluster taken from \cite{Lotz2004}.}
         \label{fig:cmd}
   \end{figure} 
In Figure \ref{fig:cmd} we show the colour-magnitude diagram (CMD) of our NGC\,1399 UCDs compared to GCs in Fornax and UCD colour data from other work. The dark blue points denote the extended UCDs with sizes above 23pc, whereas the light blue points denote all the other UCDs which are located in the three fields, for which we have FORS data. The yellow diamonds show UCDs from \cite{Evst2008}. We also include as purple triangles data for the nuclei of dwarf elliptical galaxies which are taken from \cite{Lotz2004}. The GC data (light grey dots) include the GCs of NGC\,1399, NGC\,1404 and NGC\,1389 which were taken from the Fornax ACS Survey (\citealt{Jordan2007}). The $g'$ and $z'$ AB magnitudes of the ACS Survey were transformed into the $V$ and $I$ system by using the single stellar population models from \cite{BC2003}. We assumed a GC age of 11-13 Gyrs and a Chabrier IMF with a metallicity between -2.2 dex and -0.6 dex. The transformation equation for the magnitude and colour is given as:
\begin{equation}
  V = g'-0.004-0.301\cdot(g'-z') \\      
\end{equation} 
\begin{equation}
   (V-I) = 0.445+0.518\cdot(g'-z')          
\end{equation} 
The root mean square of equation 5 is rms$=$0.010 and that of equation 6 is rms$=$0.012.
When we look at the location of the dwarf elliptical nuclei (purple) and the UCDs from \cite{Evst2008} we can see that the nuclei are on average 'bluer' than UCDs of comparable luminosity and their colours become redder with increasing nuclei luminosity. This trend towards redder colour for brighter objects is also observed for the UCDs from the literature (yellow) but with a shallower slope than for the nuclei. A large fraction of our sample of extended UCDs (dark blue) falls close to this relation, which has been noted already by \cite{Evst2008} and \cite{Brodie2011}. In contrast to their findings, we find several extended red UCDs, which are much fainter than what is expected from the colour magnitude relation established by the nuclei.

Another feature in the CMD is the position of confirmed UCDs/GCs with half-light radii $<$23 pc. Their locations in the CMD (light blue) are compatible with those of GCs (grey dots) for luminosities fainter than $M_{V}=-11$.
They also overlap in magnitude with the bright end of the GC population. In the very blue range of $(V-I)<0.9$ there is a lack of UCDs with magnitudes brighter than $M_{V}<-11$. This can be explained by the blue tilt found in various globular cluster systems (e.g. \citealt{Mieske2010,Fensch2014}). 

\begin{table*}
\caption{Results from the Sérsic fits for those 13 UCDs that have half-light radii larger than the resolution limit of $23pc$. The $(V-I)$ colours and radial velocities are taken from the respective original samples. The provided errors on $R_{eff}$ and n are those given by GALFIT and should be taken with caution, as close to the resolution limit these errors might be underestimated. The alternative object names from the literature are from the following sources: for UCDx \cite{Firth2007}, $\rm{x-xxxx}$ \cite{Mieske2002, Mieske2004}, Yxxx references \cite{Richtler2008}. Some of these objects were also presented in \cite{Richtler2005} with the following names: $^{a}$\,78:12  $^{b}$\,91:93  $^{c}$\,90:12}
\centering
\begin{tabulary}{1.0\textwidth}{ccccccccc}
\hline \hline
Name  & Name$_{\rm alt}$ & R.A. & DEC. & $V$ & R$_{\rm eff}$ & $n$ & $(V-I)$ & $v$  \\
& & (h:m:s) & ($^{\circ}$:':'')& (mag) & (pc)& & mag & (km s$^{-1}$) \\
\hline 
UCD-FORS 1 &  1\_0630 &  3:38:56.14 &  -35:24:49.0 & 20.31  $\pm$   0.004 &     28.10 $\pm$    0.16 & 3.32 $\pm$ 0.06   &    1.02 $\pm$   0.07 & 666  $\pm$ 48 \\
UCD-FORS 2 &  $\rm Y99025^{a}$  & 3:38:58.55 & -35:26:26.0 & 20.21   $\pm$   0.012   &    34.68 $\pm$     0.76 & 4.85 $\pm$ 0.17   &    0.96 $\pm$    0.04 & 1070    $\pm$ 38 \\
UCD-FORS 4 &  2\_2115  & 3:38:49.18 & -35:21:42.1 & 20.96 $\pm$ 0.003 &    23.70 $\pm$ 0.15 & 2.44  $\pm$ 0.05 & 1.19  $\pm$ 0.07 & 864  $\pm$  79  \\
UCD-FORS 32 &  UCD6  &   3:38:05.04 & -35:24:09.7 & 19.15 $\pm$   0.001     &   33.27  $\pm$   0.07 &  0.94 $\pm$  0.01     &   1.13 $\pm$       0.11 & 1220    $\pm$ 45 \\
UCD-FORS  36 &  UCDm  &  3:38:06.48 & -35:23:03.8 &   20.00   $\pm$  0.002  &     25.06  $\pm$   0.08 &   4.09$\pm$ 0.04   &     1.00 $\pm$     0.04 & 1442    $\pm$ 123 \\
UCD-FORS  45 &  UCD28  &  3:38:10.73 & -35:25:46.2 &    19.97 $\pm$   0.003   &    33.83   $\pm$   0.14 & 3.81 $\pm$  0.03     &   1.29 $\pm$    0.04 & 1715 $\pm$ 90 \\
UCD-FORS  48 &  UCD31 &  3:38:16.51 & -35:26:19.3 &    20.60 $\pm$    0.008   &    25.88   $\pm$   0.29 &  4.41 $\pm$ 0.13    &   0.95 $\pm$    0.05 & 899  $\pm$ 85 \\
UCD-FORS  50 &  1\_2095 & 3:38:33.82 & -35:25:57.0  &    20.61 $\pm$  0.007  &   24.05 $\pm$  0.24  & 4.56 $\pm$ 0.13 &   1.10 $\pm$  0.04 &   1223 $\pm$   221 \\
UCD-FORS  52 &  2\_2127 & 3:38:11.69 & -35:27:16.2 &   20.81   $\pm$  0.02   &    44.41   $\pm$    2.47 &  9.05 $\pm$ 0.38   &     1.16 $\pm$    0.06  & 1443  $\pm$ 131\\
UCD-FORS  56 &  Y10056 & 3:38:38.77 & -35:25:55.2  &    21.18  $\pm$   0.033   &    36.72  $\pm$     2.37  &  5.44 $\pm$ 0.42     &   1.26 $\pm$      0.07  & 1062  $\pm$ 105 \\
UCD-FORS  71 &  Y10048 & 3:38:35.23 & -35:25:39.2  &    22.05  $\pm$    0.015  &     24.85 $\pm$     0.51 &  2.71 $\pm$ 0.20     &   1.09 $\pm$     0.04  & 1601  $\pm$ 30 \\
UCD-FORS 81 &  $\rm UCD2^{b}$ & 3:38:06.29 & -35:28:58.8  &    19.31   $\pm$     0.048 &  37.82 $\pm$ 2.58 & 2.03 $\pm$   0.33  &    1.13 $\pm$      0.11  & 1249  $\pm$ 37 \\
UCD-FORS  84 &  $\rm 2\_2072^{c}$ & 3:38:14.69 & -35:33:40.7  &   20.77  $\pm$   0.001  &  41.85   $\pm$   0.11  &  1.96 $\pm$ 0.01 &  0.89  $\pm$   0.05    &   1448   $\pm$   176 \\
\end{tabulary}
\label{ucddat}
\end{table*}
\medskip

\subsection{Luminosity-effective-radius relation}  
\label{sec:34}
    \begin{figure*}
   \centering
   \includegraphics[width=17.5cm]{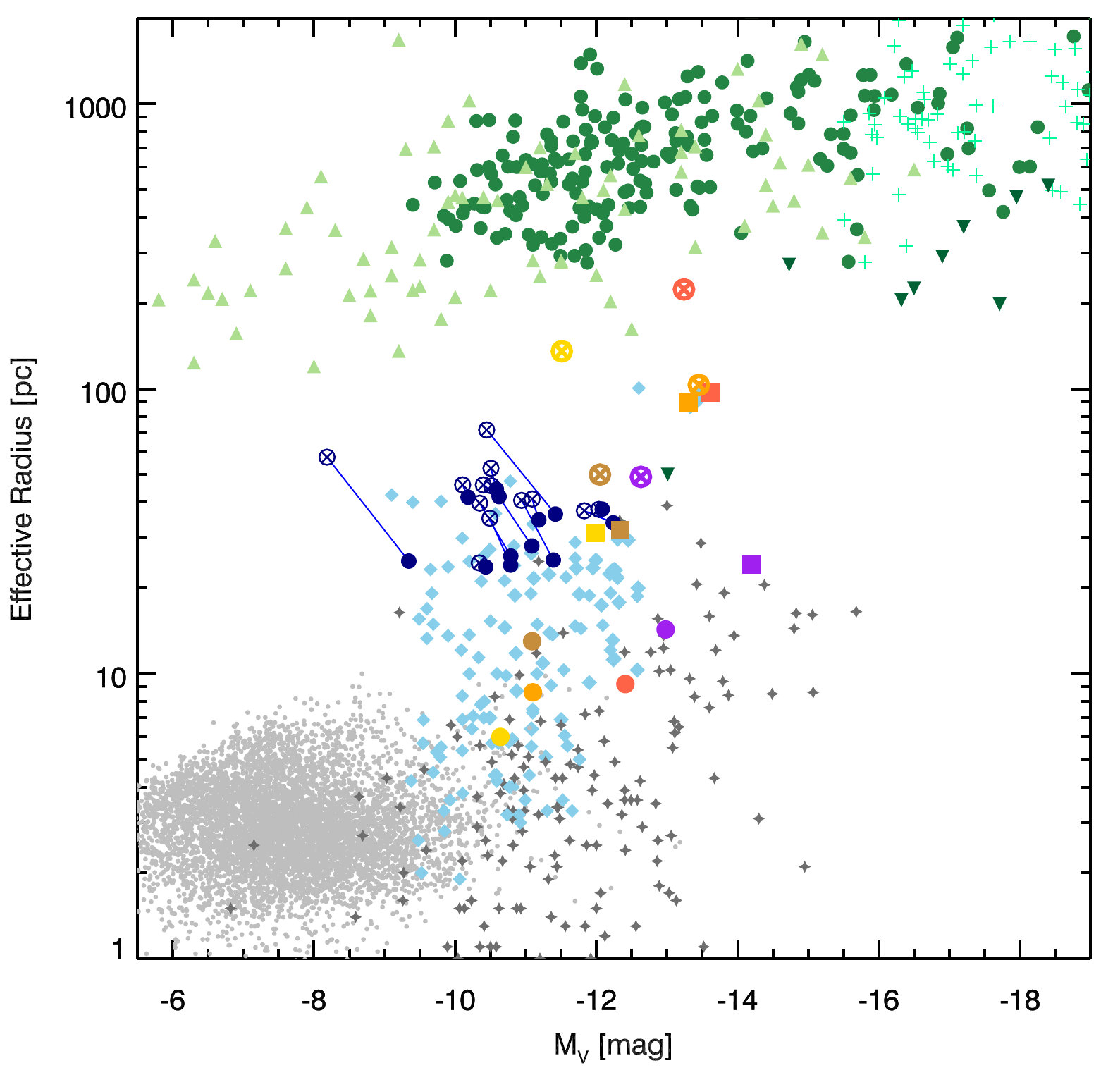}
      \caption{The relation between the effective radius and the absolute visual magnitude of several types of early-type systems are shown. As green dots we show dEs in the Hydra and Centaurus clusters taken from \citealt{Mis2008, Mis2009}. The light green triangles are dSphs in the Local Group from \cite{Mcconn2012}. The inverted green triangles are compact ellipticals from \cite{Price2009}. Light green plus signs represent early-type galaxies from \cite{Ferrarese2006}. Light grey dots are the Fornax Cluster GCs from the ACVCS survey (\citealt{Jordan2007}). The dark grey star symbols are the nuclear star clusters from \cite{Georgiev2014}. In light blue we show an assembly of UCDs with measured sizes taken from \citealt{Mieske2008, Brodie2011, Forbes2013,Evst2008, Hasegan2005}. As navy blue dots we plot the effective radii from the Sérsic profile fit. The connected navy circles with a cross shows which sizes we derived for the envelopes of a two component King+Sérsic model to each of our UCDs when we assume a 11.6\,pc King component in their centres. The yellow, orange, red and brown symbols show the 5 known decomposed UCDs with clear double profiles in the literature. Squares are the single component fits, dots the core component and circles with crosses the envelope component. The red points denote (VUCD7), orange (UCD3) and yellow (UCD5), all taken from \cite{Evst2007}. Purple symbols show M60-UCD1 taken from \cite{Strader2013} and the brown symbols are M59cO taken from \cite{Chili2008}.}
         \label{fig:sizemag}
   \end{figure*}
One of the main theories for UCD formation is that they are the isolated nuclei of larger dwarf ellipticals, being stripped of their stellar envelope through tidal interactions (e.g. \citealt{Bekki2003, Drinkwater2003}). In \cite{Pfeffer2013} the the trajectories from simulations in the luminosity-size plane of such a process were shown. During the tidal interaction dwarf galaxies lose their envelopes and, from their original position in the size-luminosity diagram as dwarf ellipticals (green points in figure \ref{fig:sizemag}), they end up with sizes and magnitudes of UCDs, effectively having to cross the empty region in between the galaxy and the star cluster branch. One could naively ask: if these objects are formed by tidal stripping -- why do we not see more objects in between both branches being currently transformed? This is due to the fact that the transformation timescale is rather short (\citealt{Pfeffer2013}) and that dwarf galaxy destruction has happened in the early phases of cluster formation, i.e. several Gyr ago (\citealt{Pfeffer2014}). This makes objects in the transition phase between the two branches very rare at present time. Additionally, if the surface brightness of the stellar envelope component reaches faint levels very quickly, we miss them in observations because there have not been systematical or deep enough searches for low surface brightness features around a large sample of UCDs yet. Nevertheless, even today galaxy clusters still accrete sub-structures and some dwarf galaxies get disrupted. Therefore, we should be able to find some "smoking gun" evidence for the tidal nature of these objects at low surface brightness levels.

In figure \ref{fig:sizemag} we plotted the effective size and magnitude of dwarf ellipticals and dwarf spheroidal (green and light green), globular clusters (light grey dots), UCDs (blue diamonds) and galaxy nuclei (dark grey diamonds). The five literature UCDs with double component profiles are shown in different colours (red, yellow, orange, brown and purple) with a circle for the core component, square for the single component and a crossed circle for the envelope. 

The structural parameters for UCDs from our sample, which could be fitted with a double profile, are shown in dark blue. The envelope component for each UCD from the two component decomposition is shown as circle with a cross. The single components are the dark blue points. Each single component UCD is connected with a line to it's derived envelope component. Wo do not show the core component since it was artificially fixed to a size of 11.6\,pc.

For the literature UCDs as well as the new sample the derived envelope parameters fall right in between the galaxy and star cluster branch. The core components from the literature UCDs all fall into a lower size UCD range close to the nuclear star cluster range. This might be a hint that UCDs with envelopes are actually transition objects in the stripping process of dwarf elliptical galaxies, or alternatively are merged super star cluster complexes with an extended envelope component, which is a result of the star cluster merging process (e.g. \citealt{Fellhauer2002}).

 \begin{table*}
\caption{Results from the double profile fits to the surface brightness profiles of the UCDs. Here we used a profile consisting of a King core fixed at 11.6pc in effective radius and a variable S\'{e}rsic Envelope component. The second and third column show the magnitude of both components. The fourth column shows the effective radius of the S\'{e}rsic envelope and the fifth column the best fit S\'{e}rsic index. The relative change in $\chi^{2}$ between the K+S fit and the original S\'{e}rsic only fit is shown in the second last column. In the last column the expected tidal radius for these objects according to equation \ref{eq:tidal} are shown. $^{1}$ For object UCD-FORS\,32 the results for a double S\'{e}rsic fit are shown}
\centering
\begin{tabulary}{1.0\textwidth}{cccccccccc}
\hline \hline
Name  & $R_{\rm King}$ & $R_{\rm Envelope}$ & $R_{\rm Envelope}$ & $n$ & Rel. Change $\chi^{2}$ & $r_{tidal}$ \\
& (mag) & (mag) & (pc)& & \% & (pc) \\
\hline 
UCD-FORS 1  & 21.28  $\pm$   0.09 &  20.88 $\pm$  0.06 & 45.67 $\pm$ 2.32   &    1.84 $\pm$   0.15 & 3.08 & 590\\
UCD-FORS 2 & 22.96   $\pm$   0.07   &    20.30 $\pm$     0.09 & 41.04 $\pm$ 6.38   &    4.76 $\pm$    0.31 &  1.57 & 444\\
UCD-FORS 4  & 23.71 $\pm$ 0.13 &    21.05 $\pm$ 0.28 & 24.45  $\pm$ 4.40 & 2.25  $\pm$ 0.19 & -0.13 & 684 \\
UCD-FORS $32^{1}$ & 19.60 $\pm$   0.04     &   20.32  $\pm$   0.02 &  88.68 $\pm$  1.39     &   0.59 $\pm$  0.03 & 44.28 & 985\\
UCD-FORS  36  &   21.17   $\pm$  0.10  &     20.45  $\pm$   0.11 &   40.57 $\pm$ 5.22   &   3.33 $\pm$ 0.30 & 4.44 & 659 \\
UCD-FORS  45  &  20.53 $\pm$   0.03   &    20.95   $\pm$   0.01 & 71.75 $\pm$  0.35     &   0.70 $\pm$    0.01 & 28.60 & 592 \\
UCD-FORS  48  &    21.79 $\pm$    0.22   &    21.04   $\pm$   0.32 &  39.64 $\pm$ 16.11    &   3.46 $\pm$    0.73 & 0.23 & 253 \\
UCD-FORS  50   &    22.22 $\pm$  0.09  &   20.89 $\pm$  0.07  & 35.12 $\pm$ 5.22 &   5.36 $\pm$  0.22 &  5.45 & 160 \\
UCD-FORS  52 &  23.82   $\pm$  0.03   &    20.88   $\pm$    0.02 &  52.59 $\pm$ 12.50   &     8.88 $\pm$    0.75  & -0.14 & 349 \\
UCD-FORS  56 &    23.72  $\pm$   0.13   &    21.29  $\pm$  0.09  &  46.02 $\pm$ 10.32     &   4.99 $\pm$  0.30  &   0.9 & 229 \\
UCD-FORS  71 &    22.50  $\pm$    0.08  &     23.21 $\pm$     0.05 &  57.50 $\pm$ 2.32     &   0.31 $\pm$     0.08  & 0.85 & 136 \\
UCD-FORS 81  &  22.68   $\pm$     0.16 &  19.36 $\pm$ 0.41 & 37.74 $\pm$   12.40  &    1.88 $\pm$  0.86  & 42.90 & 776 \\
UCD-FORS  84  &   22.61  $\pm$   0.07  &  20.99   $\pm$   0.02  &  46.02 $\pm$ 0.70 &  1.40  $\pm$   0.05    &  7.05 & 587 \\
\end{tabulary}
\label{KSdat}
\end{table*}
\medskip

\section{Tidal structures and globular clusters around UCDs}
\subsection{Tidal Structures}
In our FORS sample of 97 UCDs we investigated all of them visually for signatures of tidal tails and general asymmetric features. We looked at the original images, as well as the residual images from our best-fit GALFIT models. Underlying faint features or tails are often only well visible when subtracting the light of a symmetric profile. We classify those UCDs as having tidal features when there was a coherent tidal feature  above the background level identified. The background noise level was determined locally for each UCD, as the noise level significantly varies as function on the distance to NGC\,1399. In total we found 11 out of the 97 UCDs exhibiting such tails and features.

\begin{figure}
   \includegraphics[width=\hsize]{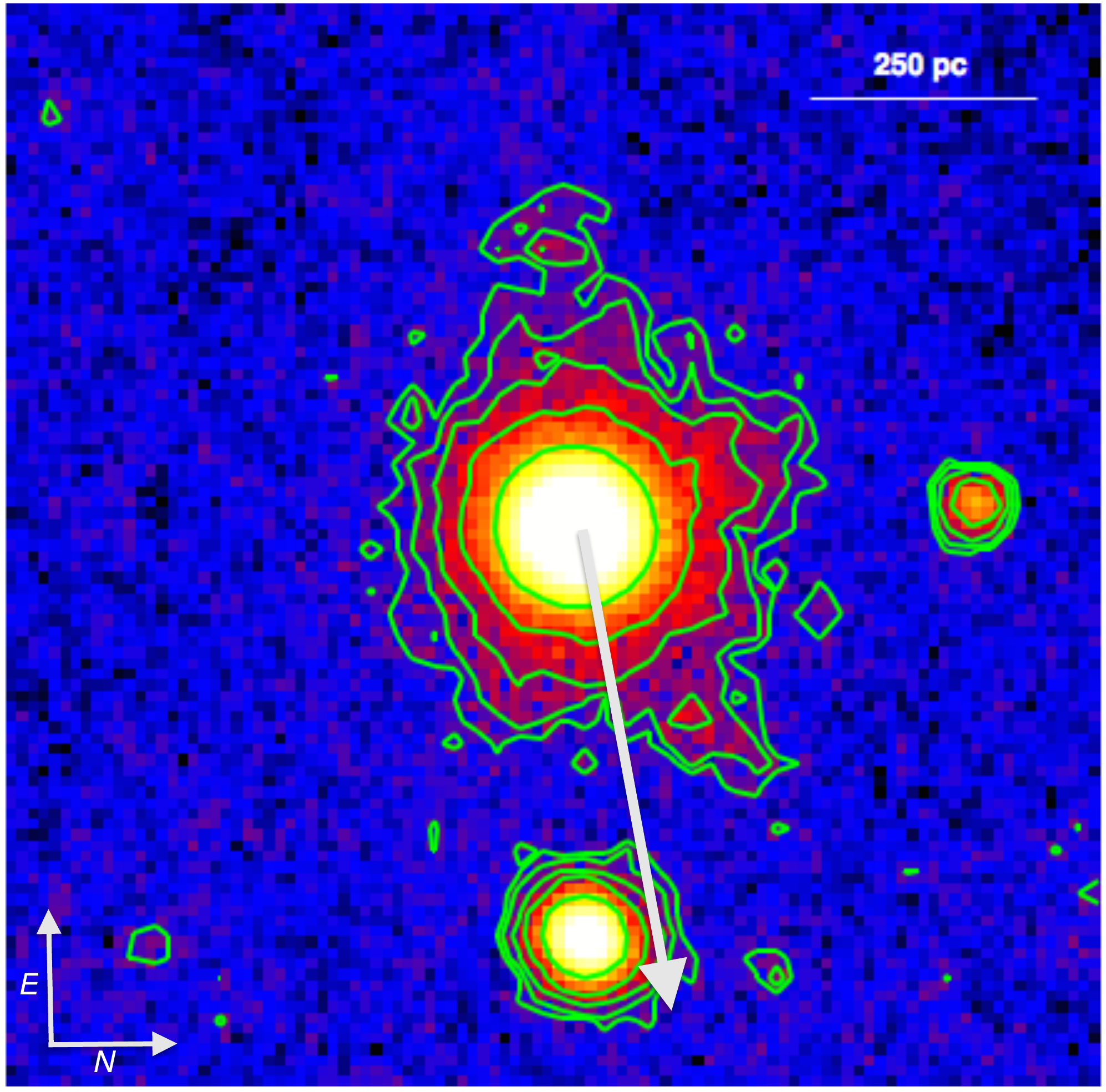}
      \caption{$R$-band image of UCD-FORS\,2 (Y99025) overlaid with green contour lines of isophotes with constant surface brightness levels. The density levels correspond to 2, 3, 5, 10 and $30\sigma$ above the background. The 2 sigma contour corresponds to a surface brightness level of $\mu=26.0$\,mag/arcsec$^{2}$. Three tidal features are clearly visible in the image. One extends towards the North-West (lower right) and is above the $3\sigma$ level for all its extension with a small $5\sigma$ peak in its middle. The tidal feature extending towards East (top) is enclosed by the $2\sigma$ isophote with a $3\sigma$ peak in the middle. On the northern side of the object (to the right) there are also clear distortions in the isophote shape visible which are around 250\,pc in extent at the 5 sigma level, and even the $10\sigma$ isophote is clearly distorted. The two large tidal tails if measured from the center of the UCD have an apparent extension of $\sim350$\,pc at $\mu=26.0$ mag/arcsec$^{2}$. The white arrow shows the direction towards the center of NGC\,1399.}
         \label{fig:nr2}
 
   \end{figure}
The most striking features we found around UCD-FORS\,2 (Y99025) as shown in figure \ref{fig:nr2}. 

We detect three clear tidal tail-like structures around this UCD. Two narrow tidal features emerge from the UCD towards the East and the North-West (NW). A smaller but brighter feature is visible on the northern side of the UCD. The NW tidal feature is above the $3\sigma$ level for mainly all its extension with a small $5\sigma$ peak in its middle. The tidal feature extending towards the East is enclosed by the $2\sigma$ isophote with a $3\sigma$ peak in the middle. These two large tidal tails, measured from the center of the UCD, have an apparent extension of $\sim350$\,pc, whereas the smaller one in northern direction is extended to $\sim250$\,pc. Anisotropies in the isophotes are detected up to $10-\sigma$. The significance of our detection combined with symmetric appearance of two collimated tails, makes it very unlikely that what we see is a projection or alignment effect. If this would be faint background features overlaid on a normal UCD we would expect much rounder isophotes at 5 and $10\sigma$ significance levels and there would be no multiple tails. In addition the isophotes on the southern side of the UCD have a much steeper profile than the ones on the northerns side which extend to larger radii at the same level.

 This UCD-FORS\,2 is also among those which have a size above the resolution limit, with a Sérsic half-light radius of $r_{\rm eff}=34.68\pm 0.76 \rm pc$, and a very blue $(V-I)$ colour of $0.96$. This object has been pointed out before by \cite{Richtler2005} for appearing to have two very faint features and potentially a faint envelope around it.They also found strong Balmer lines, which usually indicates the presence of a young stellar population. The alternative explanation of a very low metallicity is not supported because of its Washington colour that points to a metallicity of $>-1.3$dex.

    \begin{figure}
   \centering
   \includegraphics[width=\hsize]{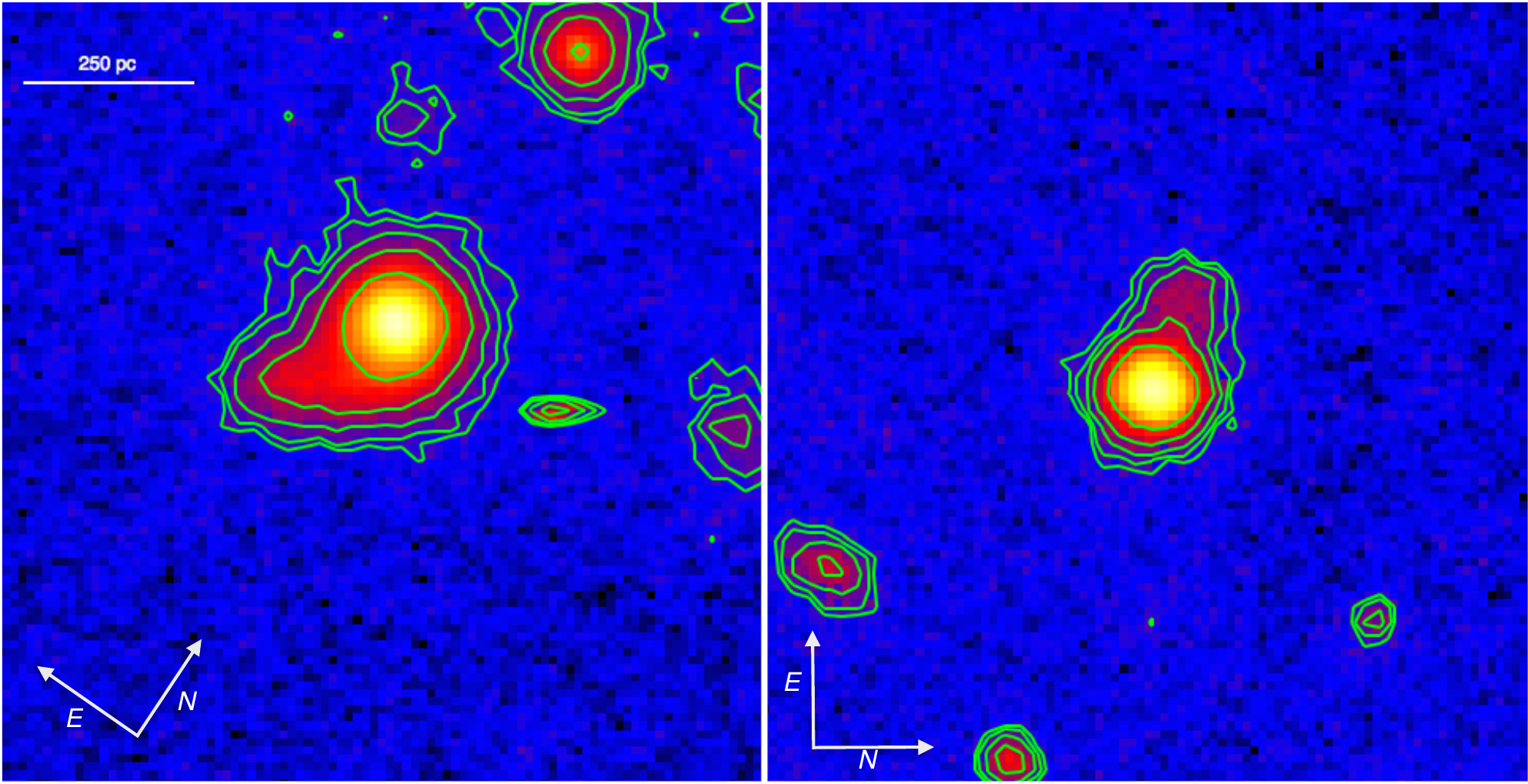}
   \caption{The left image shows the $R$-band image of UCD-FORS\,94. In the right panel UCD-FORS\,7 is shown. Both are overlaid with green isophote contours of constant surface brightness levels. The density levels correspond to 2, 3, 5, 10 and $30\sigma$ with the lowest contour corresponding to a surface brightness level of $\mu=26.01$\,mag/arcsec$^2$. }
         \label{fig:twotails}
   \end{figure}

In figure \ref{fig:twotails} on the left hand side the $R$-band image of UCD-FORS\,94 is shown, on the right panel UCD-FORS\,7 is displayed. Overlaid on both images are the contour levels of constant surface brightness as green lines. We display the 2, 3, 5, 10 and 30$\sigma$ levels with respect to the background noise level for each image. The UCD in the left panel shows a strongly pronounced and extended feature towards South-East. Compared to the tails of UCD-FORS\,2 of image \ref{fig:nr2} this is a much brighter feature. Already the $10\sigma$ contour level is clearly shaped like a tail-like extension. Measuring the full extent of the $5\sigma$ contour level of this feature, from the center of the UCD outwards, gives us an apparent size of 230\,pc. The surface brightness level of this extension is higher than expected for a tidal tail. It is possible that we see a background object in projection. Spectroscopy or high resolution imaging are needed to clarify the nature of this apparent UCD extension.

The feature we detect at UCD-FORS\,7 towards the North-East has a much lower surface brightness level and a smaller extension than that of UCD-FORS\,94. The full extent of the $5\sigma$ contour along the tail-like feature is 175\,pc. This might be a true tidal tail but spectroscopic confirmation is needed for this target as well.

In the following we estimate the probability of by-chance superpositions of low surface brightnes background galaxies on our UCD sample.
For simplicity we assume a poisson distribution of UCDs and background galaxies. Within the surface area of our FORS fields we have $N_{1}=97$ confirmed UCDs. We determined the amount of extended galaxies that could be confused as tidal tails if perfectly aligned by running SExtractor \citep{Bertin1996} on our FORS observations. We restricted the SExtractor output to objects with an ellipticity ($\epsilon=1-\frac{B}{A}$) of $0.2<\epsilon<1.0$ to retain only objects that have a significant elongation. We also set the requirement that the CLASS\_STAR (CS) parameter is $CS<0.5$ to only get objects that have been classified as galaxy-like objects. We set the detection threshold to $2-\sigma$ above the background with a minimum of 10 contiguous pixels above this threshold. In total we extracted 1205 objects which could mimic a tidal tail when projected on a UCD-like object. As a minimum distance within which such a projection might be confused with a tidal tail we assume $r=2\arcsec=0.184\rm kpc$. The combined surface of our FORS fields is $\Omega=35.595\cdot 10^{3} \rm kpc^{2}$. The probability to find one such close overlap in our sample is given as:

\begin{equation}
 P(R<r)=N_{1}\cdot N_{2} \frac{\pi \cdot r^{2}}{\Omega}=0.349
\end{equation}
Thus, the probability of 34.9\% for one superposition is not negligible. However, the probability that all 11 detected tail candidates are random superpositions is much lower. Calculating the binomial probability for two or more random superpositions within our sample of 97 UCDs is already P=5.09\%. For three and respectively four or more random overlaps among our 11 candidates the probability goes down to P=0.53\% and 0.04\% respectively. The probability that all 11 detections among the 97 UCDs are random overlaps comes out as P=$7.25\cdot 10^{-14}$. Thus it is virtually impossible that all detected tails are just random overlaps and it is very likely that the majority of our objects are detections of true tails around UCDs.

\subsection{Globular clusters around UCDs}
\label{sec:companion}
Another feature we detected frequently in the images of our UCDs are faint 
GC candidates,

 which are located very close to the main object. In the following we investigate the frequency of close companions within 300\,pc radius around each UCD. This radial size is motivated by the tidal radius of a $M=10^{6}M_{\odot}$ UCD at a pericenter distance of 30\,kpc from a host galaxy with $M= 10^{12}M_{\odot}$:
\begin{equation}
  \label{eq:tidal}
  r_{tidal}=R\cdot \left(\frac{m}{M}\right)^{\frac{1}{3}}
\end{equation}
In total, we found 19 UCDs with a faint companion source that is closer than 300\,pc in projection. This is $\sim$20\% of our total UCD sample. 
Notably, four of these objects with companions belong to those with extended sizes above the resolution limit of 23\,pc (UCDs in panels a), d), g), q) in figure \ref{fig:extended}). For some cases the object is so close that it blends into the surface brightness profile of the main UCD, which makes the object appear skewed and elongated (see e.g. UCDs in the panels a), c), m), n) and p in figure \ref{fig:extended}). Fitting a symmetric profile to these objects and then investigating the residuals shows that there is clearly a faint point source underneath and that the UCD is not just elongated in one direction. The second type are those UCDs where the faint source is well separated from the UCD and the two sources are clearly distinguishable.

    \begin{figure}
   \centering
   \includegraphics[width=\hsize]{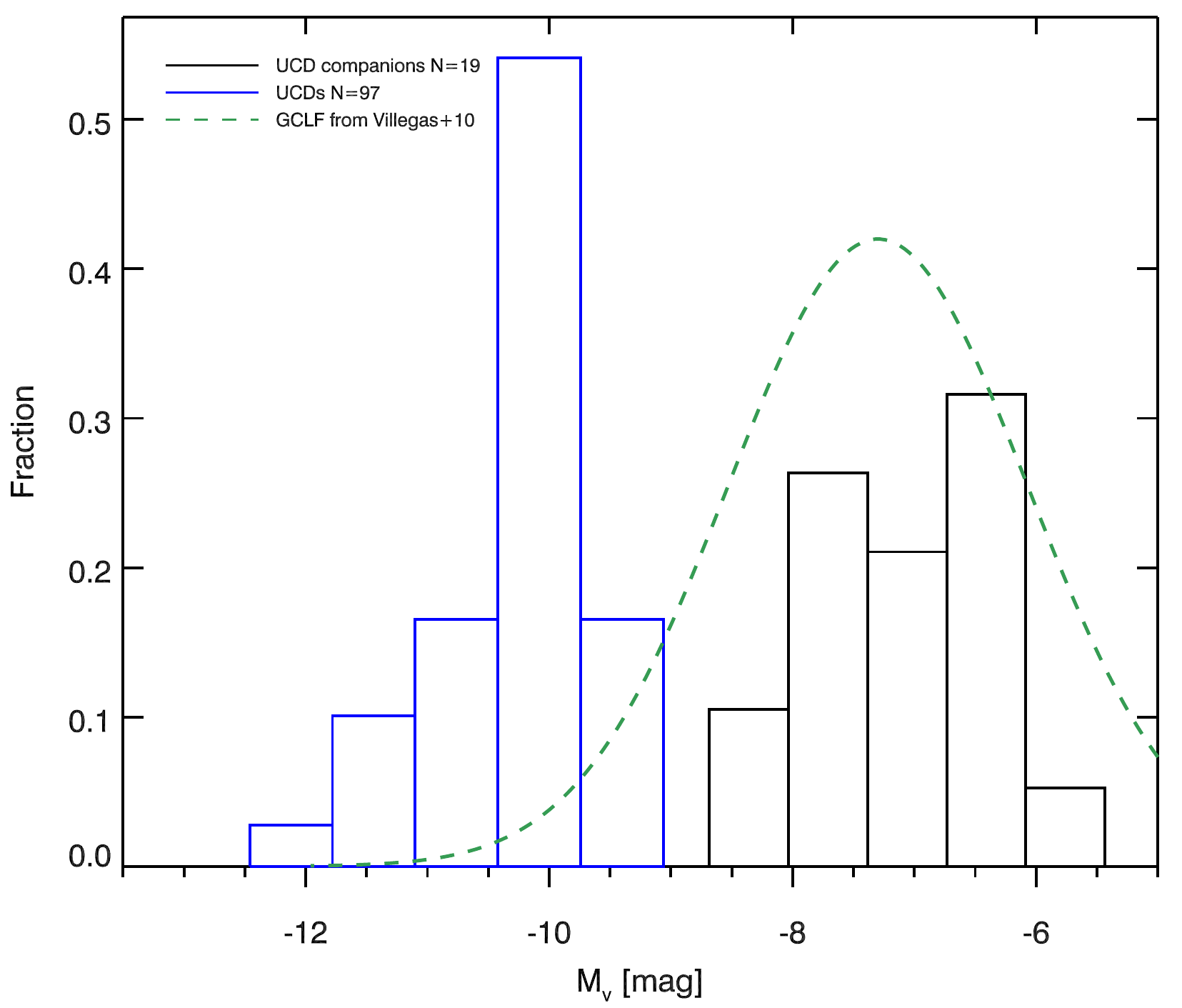}
      \caption{In blue the histogram of absolute magnitudes of the 97 UCDs in the FORS field is shown. In black the histogram of the 19 point sources we detected within a radius of 300\,pc around the UCDs is shown, if we assume they lie at the distance of the Fornax cluster. The dashed green line is a Gaussian with peak magnitude $M_{V}=-7.3$ mag and a dispersion $\sigma=1.23$\,mag, as measured for the globular cluster luminosity function of NGC\,1399 in \cite{Villegas2010}. The NGC\,1399 GCLF is normalized to the number counts of the companions in the magnitude bin at $M_{V}\simeq-6.5$ mag. The average luminosity of the measured companion sources is $M_{V}=-7.04$}.
         \label{fig:histmag}
   \end{figure}
All UCDs from our sample have radial velocities that confirm them as Fornax cluster members. As there are no measurements of radial velocities of their faint companion sources we take a statistical approach to determine if these are potential globular clusters that are associated with the UCDs. For this we measured the magnitude of the point sources on the residual images, where the light of the main UCD was subtracted. Then we assumed that these objects lie at the distance of Fornax at $(m-M)=31.39$. In figure \ref{fig:histmag} the histogram of the absolute magnitude distribution of the UCDs is shown in blue and their companion sources in black. The distribution of companion sources peaks at $M_{V}\simeq-7.0$. Overplotted as green dashed line is the Gaussian fit to globular cluster luminosity function of GCs around NGC\,1399, as derived by \cite{Villegas2010}. The Gaussian has a peak magnitude of $M_{V}=-7.3$ and a dispersion of $\sigma=1.23$.

All properties of the UCDs and their companions are given in table \ref{companiondat} in the appendix. We also estimated the tidal radius for each UCD in this sample (column $r_{\rm tidal}$ in table \ref{companiondat}) based on their projected distance from the center of  NGC\,1399 and their magnitudes, with equation \ref{eq:tidal} given above. We then checked if our possible companions would lie within the tidal radius of their host UCD. For that we took the ratio between $dist_{\rm comp}/r_{\rm tidal}$ (see column 9 in table \ref{companiondat}) which denotes at which fraction of the tidal radius of this object the companion is located. For 16 out of 19 objects this fraction is below 1.0 which means the companion candidates are well within the tidal radius. Only for 3 objects (b, j, l) the companions are actually outside of the predicted tidal radius. Thus if these companions are associated with the UCD it is likely that they are still bound to it.

We also checked if any of the companion GC candidates was detected by the \cite{Dirsch2003} observations. We found that the GC candidate from panel b) at the bottom of the frame was indeed detected and has a colour of $C-T_{1}=1.60$. In the colour magnitude space this puts it at the border between ``blue" and ``red" GCs at $C-T_{1}=1.55$.   
\begin{figure}

   \centering
   \includegraphics[width=\hsize]{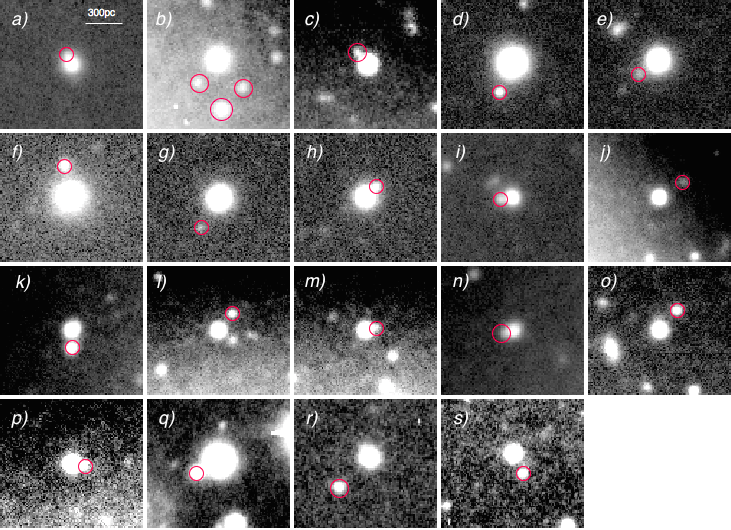}
      \caption{Cutout images of the 19 UCDs which have companion point sources (red circles) within a radius of r$<$300\,pc. As the cutout images have varying background levels, the contrast used for display can slightly vary between the images to enhance the visibility of the companion.}
         \label{fig:comp}
   \end{figure}

\section{Spatial clustering of GCs around UCDs in the halo of NGC\,1399}
\subsection{The globular cluster system around NGC\,1399}
To study the statistics of globular cluster clustering around UCDs we use the catalogue of globular clusters around NGC\,1399 by \cite{Dirsch2003}. The catalogue contains $10457$ point sources. It was obtained using wide-field Washington photometry. We defined GCs as point sources in the colour range $1.0<C-T_{1}<2.3$. To avoid any incompleteness we further restricted the GC sample to point sources of $T_{1}<23.5$, for which the data are still complete. Since the completeness depends on the object's colour, an additional colour dependent magnitude cut was applied similar to the original work, given as:
\begin{equation}
T_{cut} < -0.935\cdot (C-T_{1})+ 24.69
\end{equation}
In order to avoid any duplications with the UCDs we exclude all objects that have masses above $M=10^{6}M_{\odot}$. For the mass calculation of the GC candidates we applied the same simple stellar population models (SSP) from \cite{Maraston2005} as used in section \ref{sec:surf} for the UCDs to their Washington photometry, assuming an age of 13 Gyr, a \cite{Kroupa2001} initial mass function (IMF) and a red horizontal branch.

Applying all selection criteria leaves us with a total of $2884$ globular cluster candidates. In figure \ref{fig:pos2} we show the central area of the Fornax cluster with the spatial distribution of these GCs (black dots) and the UCDs (red), respectively.

For our statistical analysis on the spatial distribution of the GCs we need to take into account the incomplete areas of the wide-field imaging. In figure \ref{fig:pos2} the borders of incomplete areas and gap regions are shown as dashed lines. For any statistical analysis we have created a spatial incompleteness mask that determines which fraction of any given surface area is in the incomplete area. Thus we can statistically correct the number of GCs per surface unit. The green areas around the center of NGC\,1399 are also masked out because of the incompleteness of the GC sample so close to the center of the main galaxy. For the two neighbouring galaxies NGC\,1404 and NGC\,1387 the green masking boxes are chosen generously, to avoid any contamination of the radial distribution of NGC\,1399 GCs by the GC populations of these neighbouring galaxies. 

The final incompleteness mask combined with the dataset makes it possible to apply very accurate geometrical incompleteness corrections to the number of objects contained in each annulus.

\begin{figure}
   \centering
   \includegraphics[width=\hsize]{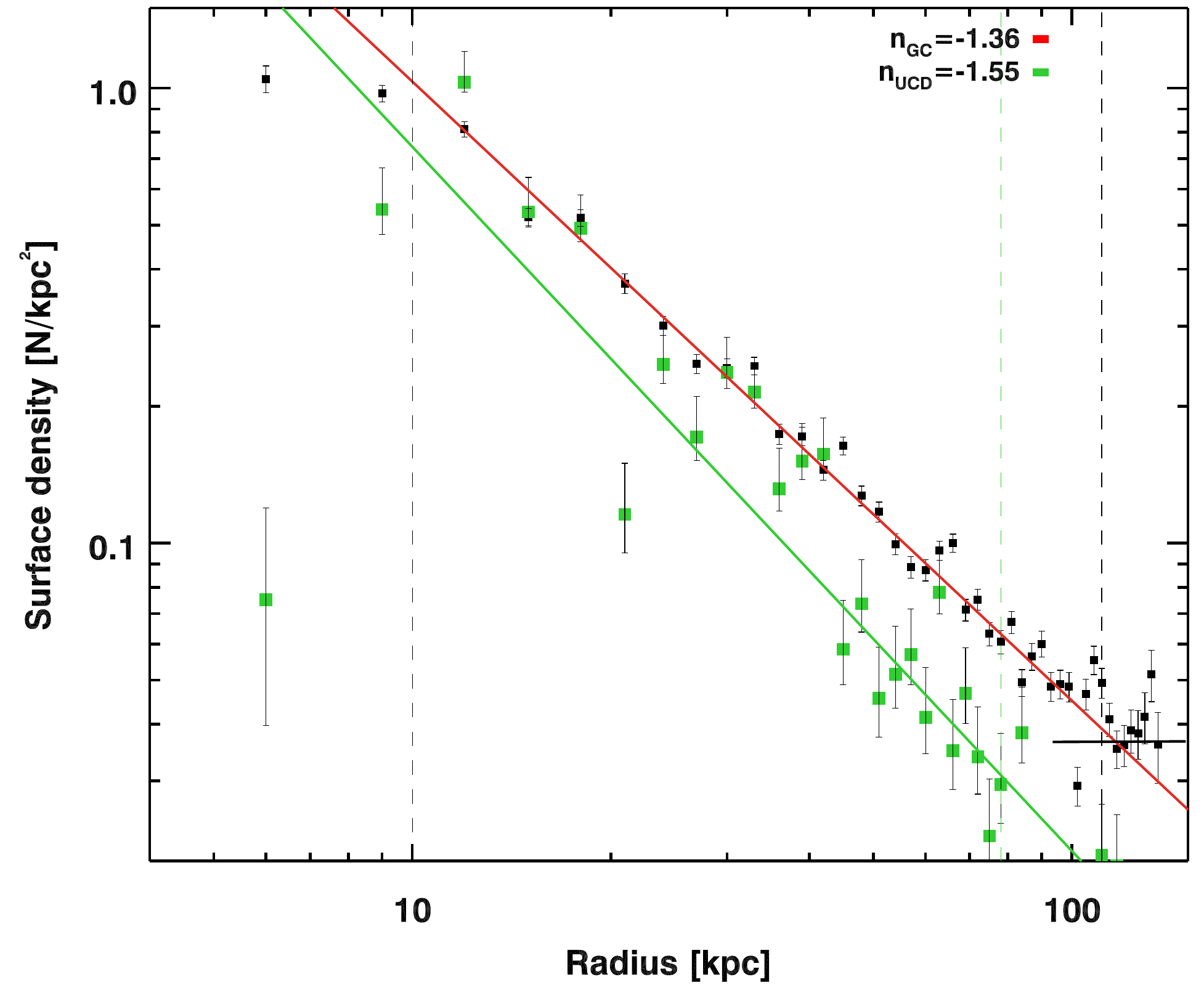}
      \caption{Projected surface density distribution of all selected globular clusters as function of their galactocentric distance to NGC\,1399. The red line shows the best powerlaw fit with an exponent of $n=-1.36$. The background level was marked with a black horizontal line. The vertical dashed black lines show the radial interval to which we restricted our fit. In green the radial distribution of the confirmed UCDs is shown. As their absolute radial density is an order of magnitude smaller than the GCs, their density is scaled up by a factor of 10 for better visibility in the plot.}
         \label{fig:gc1399}
   \end{figure}

We analyse the projected surface number density of globular clusters (black dots) around NGC\,1399 as function of their galactocentric distance to NGC\,1399. For that we adopt bins between 10 and 110\,kpc with a spacing of 3\,kpc. At radial distances $r<10$\,kpc the density profile of the GCs flattens out (see black points in figure \ref{fig:gc1399}) due to incompleteness caused by the very bright center of NGC\,1399. At distances of $r>110$\,kpc the number counts are too low and there is contamination by background objects and GCs of neighbouring galaxies. The number of GCs in each bin was corrected for geometrical incompleteness as explained above. The completeness corrected number of GCs is then divided by the surface area in physical units of $\rm kpc^{2}$ to obtain the projected surface density $\Sigma(r)$ of globular clusters, which is plotted in figure \ref{fig:gc1399}. The same procedure has also been done for the UCD sample located in the same wide-field and is shown as blue data points in figure \ref{fig:gc1399}. As the absolute surface density of UCDs is an order of magnitude smaller, the absolute density values of the UCDs (green) in figure \ref{fig:gc1399} were scaled by a factor 10. This makes it easier to compare the slopes of both populations.

The radial surface density distribution of GCs was then fitted with a power law, given as: $\Sigma(R) = (R/a_{0})^{n}$, where $\Sigma$ is the number of GCs per square kpc and $R$ is the radial distance in kpc. The fit was restricted to $10<R<110$\,kpc in galactocentric distance as shown by the two dashed vertical lines in figure \ref{fig:gc1399}. 

The resulting powerlaw fit is shown in figure \ref{fig:gc1399} as red line with the measured density values shown as black dots. The central value of our power law fit is $a_{0}=10.25$ whereas the index we derive is $n=-1.36$, which is shallower than the $n=-1.61$ derived by \cite{Bassino2006}. This is not surprising since \cite{Bassino2006} defined the slope after subtracting background counts. 

\begin{figure}
   \centering
   \includegraphics[width=\hsize]{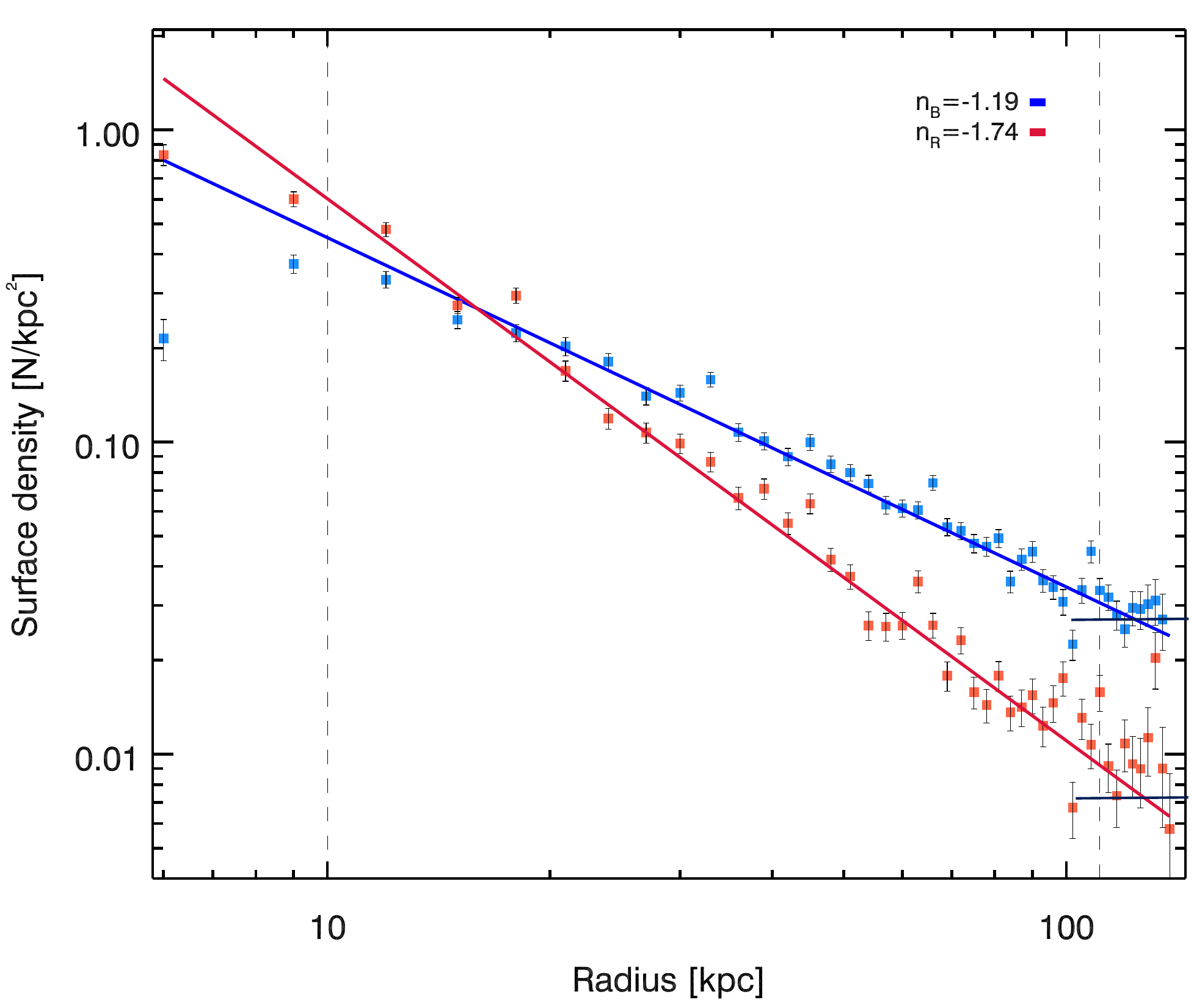}
      \caption{Projected surface density distribution of all selected globular clusters as function of their galactocentric distance to NGC\,1399. The red line shows the best fit powerlaw to the red globular clusters with an exponent of $n=-1.74$ whereas the blue line shows the best fit powerlaw to the blue GC population which has a more shallow slope of $n=-1.12$. The two short horizontal lines show the respective background levels used for the statistical correction of our signal. The dashed black lines show the radial interval to which we restricted our fit.}
         \label{fig:gc_colour}
\end{figure}

We also derived the radial surface density distribution for the blue $1.0<C-T_{1}<1.55$ and red $1.55<C-T_{1}<2.3$ GCs of our sample as shown in figure \ref{fig:gc_colour}. The squares in red and blue, respectively, show the measured surface density for each radial bin with their corresponding poisson errors. We choose $C-T_{1}=1.55$ as limit for splitting the GCs into a red and blue subpopulation, as \cite{Bassino2006} have shown that there is a dip in the bimodal colour distribution at this colour.

It is clearly visible from figure \ref{fig:gc_colour} that the red GC population has a steeper density profile than the blue GCs. The slope of the red GCs is $n_{\rm R}=-1.74$ whereas the blue GCs follow a powerlaw with $n_{\rm B}=-1.19$. The shallow profile of blue GCs and the stronger radial concentration of the red GC population in NGC\,1399 is in agreement with what has been shown in \cite{Bassino2006}. 

\subsection{Spatial correlation of GCs with UCDs}
In this section we explain the search for spatial correlations of GCs around UCDs. A positive correlation might be expected if UCDs originate either from stripped nuclei of galaxies with their own globular cluster systems of from merged star cluster complexes with remnant clusters around them. 

The stripped nuclei scenario is a viable formation channel in galaxy clusters as has been shown by \cite{Pfeffer2014}. The reasoning for finding correlation signatures is the following: luminous dwarf galaxies, and in particular nucleated dwarf galaxies, are known to host their own GC system (e.g. \citealt{Lotz2004,Georgiev2009}). When a dwarf galaxy falls into a large cluster of galaxies it starts to experience tidal forces and its tidal radius shrinks.

Through the tidal interaction with the cluster, the GCs residing outside the shrinking tidal radius are lost and disperse into the general GC population of the central cluster galaxy, whereas GCs inside the tidal radius can remain bound to the core of the dwarf galaxy for a long time (\citealt{Muzzio1986}).

One of the scenarios suggested for the formation of nuclear star clusters (NSCs) is that massive globular clusters merge towards the center of their host galaxy via dynamical friction in less than a Hubble time. (\citealt{Tremaine1975, Capuzzo1993}). This process is especially efficient in dwarf galaxies. Recently, \cite{Arca-Sedda2014} have computed, in a statistical approach, the number of surviving clusters around a galaxy and their mean mass, after a full Hubble time of dynamical friction. For dwarf galaxies with $M=10^{9}M_{\odot}$ one expects 80\% of the original globular cluster population to have merged with the nucleus via dynamical friction, and the mean mass of the remaining GC population is $M=10^{5}M_{\odot}$. 

Recent mergers of star clusters in star cluster complexes can host a variety of substructure in the form of faint envelopes, tidal tails and non-merged companion GCs (eg. \citealt{Bruens2011, Fellhauer2005, Bruens2009}). Whereas in some simulations of UCD-like objects the surrounding GC companions have merged into the central object within the first Gyr (\citealt{Bruens2011}), and thus the merging process has finished, other simulations of lower mass star cluster complexes suggest that some member star clusters still can be identified as substructures up to 5\,Gyr after the onset of merging. Thus, depending on the initial conditions of the merging star cluster complex, we might expect to find companions around it for several Gyr after its formation.

In the following we test if there is a statistical overabundance of globular clusters in close proximity of UCDs, as we expect such a signal for the inwards migration of GCs within a disrupting nucleated dwarf galaxy as well as for a merged super star cluster complex. The main discriminator between both cases is the age of the UCD, as substructure in star cluster complexes has only been shown for ages of 5\,Gyrs.

We can use the radial GC density of NGC\,1399 which we determined in the previous section as proxy for the GC density we expect at any given radial distance to the main galaxy. We then determine the local GC density around the UCDs and compare it with the expected value from the global distribution. This derived density ratio between expected and measured density shows if we have any significant clustering of GCs around UCDs for single objects and the full sample.

   \begin{figure}
   \centering
   \includegraphics[width=\hsize]{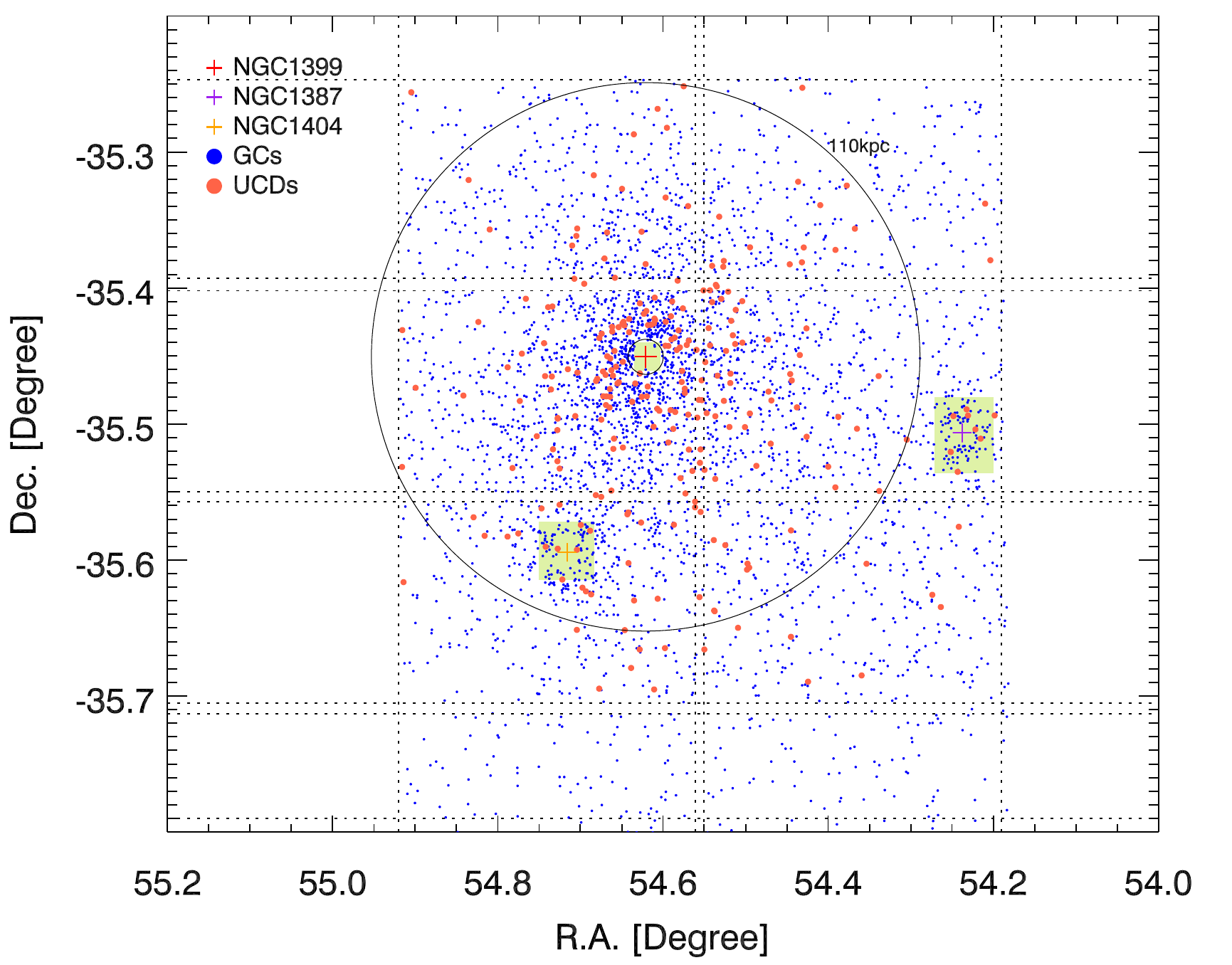}
      \caption{The central area of the Fornax cluster is shown. The globular clusters are denoted as blue dots, UCDs as red dots and NGC\,1399 as well as two neighbouring galaxies are shown as coloured plus signs. The dashed black lines mark the borders of the CCD chips and gap areas of the globular cluster sample. For our statistical studies these will be masked out. The green areas are the ones that will also be masked out in the final study because they host their own GC population. The small green circle locates the position of the UCD with 4 neighbouring GC candidates which is shown in figure \ref{fig:cand} in close up.}
         \label{fig:pos2}
   \end{figure}
In our statistical study we include 206 UCDs from the original sample, which lie within the Washington photometry area and are more massive than $M> 10^{6}M_{\odot}$. Their spatial distribution is compared to the 2884 GC candidates from the Washington photometry sample which have masses less than $M< 10^{6}M_{\odot}$ and also fulfill the colour selection criteria for GCs (see previous section). The green areas in figure \ref{fig:pos2} are excluded for being too close to a bright center of a galaxy and the chip gap areas, indicated by dashed black lines, are also excluded because. 
 
To determine the local density of GCs around UCDs we used the same method as for the large scale distribution. In ring shaped apertures around the UCDs we determined the GC surface density for 8 radial bins from 0.5\,kpc up to 4\,kpc with a spacing of 0.5\,kpc. The surface area of each annulus is completeness corrected if a part of it lies in the area that is masked out. The measured surface density $\Sigma_{\rm measured}$ of GCs around each UCD in each radial bin is then compared to the density we would expect at this position of the UCD from the derived large scale GC distribution (see figure \ref{fig:gc1399} and \ref{fig:gc_colour}). The expected and measured GC density distribution are both statistically background corrected before the clustering signal is calculated. For the full population and the red and blue subpopulation we derive the following background GC densities from figure \ref{fig:gc_colour}: $\Sigma_{0}(\rm All)=0.026kpc^{-2}$, $\Sigma_{0}(\rm Blue)=0.026kpc^{-2}$ and $\Sigma_{0}(\rm Red)=0.0073kpc^{-2}$.

After deriving the clustering ratio $\Sigma_{\rm measured}/\Sigma_{\rm expected}$ for each \textit{single} UCD we average the clustering ratios for our full sample at each radial bin. The errorbars refer to the standard deviation of the clustering signals, divided by the square root of the number of UCDs we have averaged. The same statistical procedure was applied to the blue and red GC subpopulations separately, where the expected GC surface density was taken from the colour separated profiles shown in figure \ref{fig:gc_colour}.

The two black circles in figure \ref{fig:pos2} show the area between 10 and 110\,kpc of galactocentric distance to which we restrict the positions of the UCDs for which we do the statistical clustering study. This is the radial range within which the measured radial GC density profiles do not deviate much from the power law (see figure \ref{fig:gc1399}). At larger radial distances the noise in the surface density of the global distribution is too high and in the central areas below 10\,kpc the surface density profile flattens out, due to observational incompleteness caused mainly by the bright central parts of NGC\,1399.

\begin{table*}
\caption{Summary for the clustering signals of GCs around UCDs, for the 100 UCDs which have a GC candidate within 1\,kpc. The clustering check has also been subdivided into just looking at the red and blue GCs.} 
\centering
\begin{tabulary}{1.0\textwidth}{ccccccc}
\hline \hline
Radial bin [kpc] & $C$ & $\sigma$ & $C_{\rm red}$ & $\sigma_{\rm red}$ & $C_{\rm blue}$ & $\sigma_{\rm blue}$ \\
\hline 
0.5  &    3.60 $\pm$   0.70  & 3.70  &  2.62  $\pm$   1.62   &   0.99 & 4.64  $\pm$   1.08   &   3.36 \\
1.0  &    3.79 $\pm$    0.76  &  3.69  &    2.97  $\pm$  1.50  &   1.31 & 5.27  $\pm$   1.46  &    2.92\\
1.5  &   0.92 $\pm$  0.15  &  -0.53  & 0.43  $\pm$   0.46   &   -1.24 & 0.97 $\pm$  0.23  &  -0.13 \\
2.0  &   0.96  $\pm$  0.19  &  -0.18   &     0.59  $\pm$   0.49   &  -0.84 & 1.05  $\pm$  0.28  &   0.16 \\
2.5  &   0.64 $\pm$   0.10  &  -3.31   &   0.22  $\pm$ 0.44 &    -1.75 & 0.67  $\pm$  0.16 &  -2.09 \\
3.0  &   1.19 $\pm$  0.43  &  -0.45  &    0.55  $\pm$  0.48  &  -0.95 &  1.64 $\pm$  0.90  &  0.71 \\
3.5  &   0.94  $\pm$  0.15 &  -0.38 &  1.32  $\pm$  1.01  &   0.31  &  0.96  $\pm$  0.16  &   -0.25 \\
4.0  &   0.98  $\pm$ 0.10 &  -0.15 &  6.54  $\pm$  5.37  &   1.03 &  0.96  $\pm$  0.16  &  -0.26 \\
\end{tabulary}
\label{cluster}
\end{table*}
     \begin{figure}
   \centering
   \includegraphics[width=\hsize]{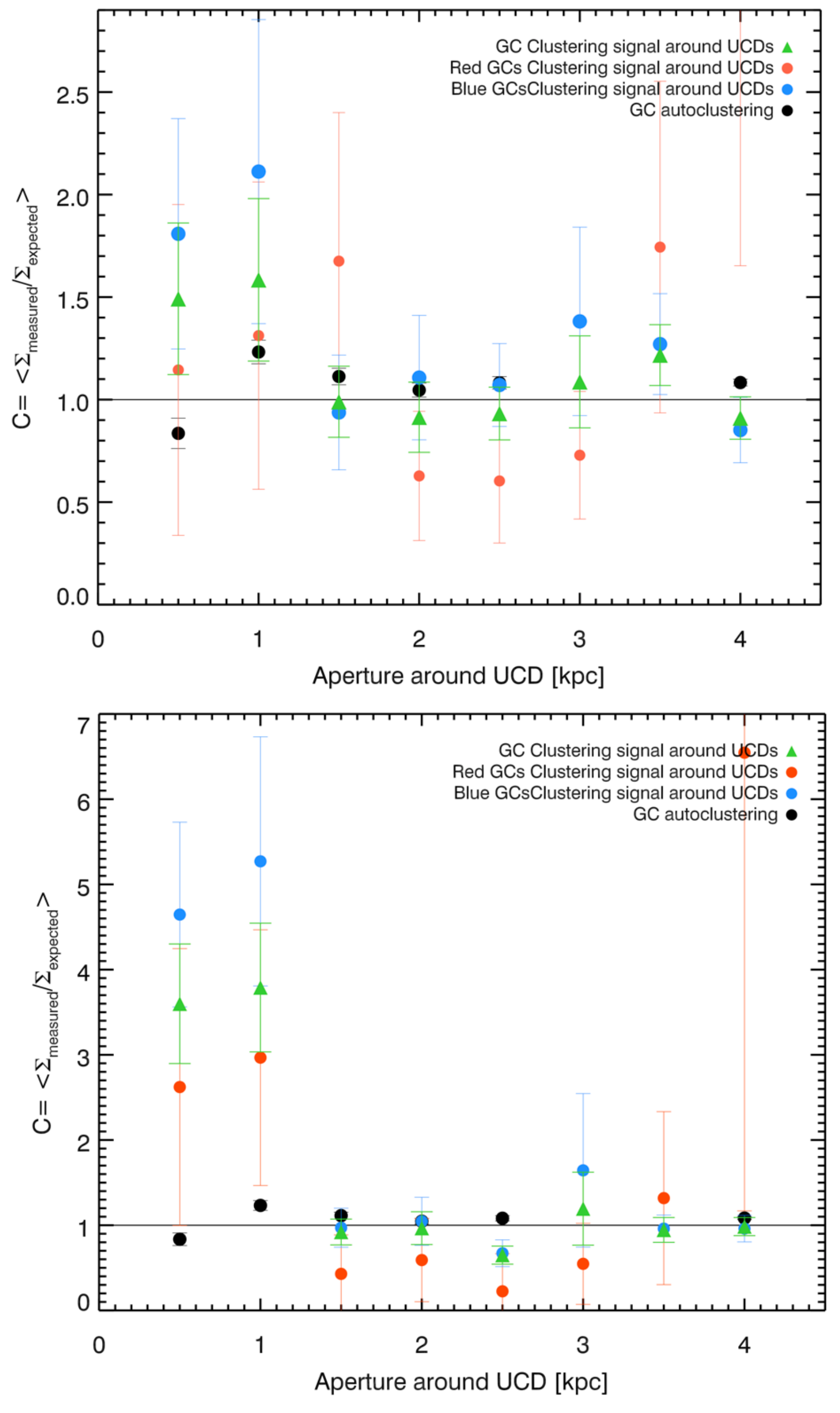}
      \caption{In the top panel the average clustering of GCs around UCDs at different radial bins from the UCDs for the full GC population (green) and the red and blue subpopulations are shown. The black data points are the comparison with the autoclustering of the GCs with themselves as a baseline. The clustering signal at each radial bin is the ratio between measured surface density and expected surface density of globular clusters from the large scale distribution, at different radial distances from the UCDs. The black line denotes the null level at which we have no clustering signal compared to what we expect. We averaged the clustering signal for each respective radial bin over the full probed sample of 206 UCDs. Bottom panel: Here we show the same plot for the 100 UCDs for which we found a GC candidate within $r<1$\,kpc. Note the different scaling of the two y-axis. }
         \label{fig:clust}
   \end{figure}

In the top panel of figure \ref{fig:clust} the clustering results are shown for all the 206 UCDs that we included in the statistical sample. As we do not expect all UCDs to have a remnant GC nearby we also studied the subset of UCDs that have an object within 1\,kpc. In total 100 UCDs, which is close to half of the studied sample, have a GC candidate within 1\,kpc of radius. The clustering results of those UCDs are shown in the lower panel of figure \ref{fig:clust}.
In both panels of this figure, the \textit{average} clustering signal $C$ for the full GC population is shown as green triangles, and the results for red and blue GC subpopulations in their respective colors. The numeric results of the bottom panel are also summarized in table \ref{cluster}, showing the average clustering signal for each population with their errorbars, as well as the sigma significance of this signal. As a comparison, the spatial autoclustering of the 2884 GCs is shown as black datapoints in figure \ref{fig:clust}. 

For the clustering signal of all UCDs we find an increasing overdensity of GCs towards smaller distances below 1\,kpc. At the smallest radial bin we find that the average UCD has 1.5 times more GCs within 500\,pc than we would expect from the underlying distribution. Due to the high scatter in this average this results in a 2$\sigma$ statistical significance for the smallest bin. The color separated clustering signals show that for the red GCs $C$ is compatible with 1, which implies no particular overabundance with respect to the expected values. The high scatter is an indication though that, although some individual UCDs have clustered red GCs, this is not the case on average. The blue (metal-poor) GCs show a higher abundance around UCDs than their red counterparts in the two inner bins. The statistical significance at 1.0\,kpc is 1.5$\sigma$.

Looking at the clustering signal of the 100 UCDs with a GC candidate within 1\,kpc (bottom panel of figure \ref{fig:clust}), we see a clear clustering signal that distinguishes this UCD population from the autoclustering signal of GCs (black dots) at a very high significance.
The clearest clustering signals we find for the 0.5 and 1\,kpc radial bins of this sample. The detected overabundance of GCs around these UCDs has a statistical significance of 3.7$\sigma$ 
for the two smallest bins (see green triangles in bottom panel of figure \ref{fig:clust} and table \ref{cluster}). This clustering signal increases for the blue GCs and decreases for the red GCs, if we compare the clustering of the red and blue subpopulations separately. The blue subpopulation shows an even higher clustering of $C=4.64$ at 0.5\,kpc compared to the red population with $C=2.62$. In other words, blue, and thus most probably metal-poor GCs, are on average 4.6 times more abundant within 0.5\,kpc of a UCD than what we would expect if the blue GCs were randomly distributed. Also, for the blue GCs the significance of the clustering is at $3.36\sigma$. Red GCs are on average also more abundant around UCDs up to 1\,kpc but only half as much as the blue GCs, and due to the small number statistics for the red subpopulation their numbers have significantly larger errors and much lower sigma values. All the statistics for the bottom panel are also summarized in table \ref{cluster}. In the radial bins larger than 1\,kpc we do not find any significant clustering signals. 

We note that this is an average clustering value for all UCDs. Single objects can be still located within local over- or under-densities of GCs but if the \textit{averaged} ratio is significantly higher than 1.0 it means that there is a systematic deviation of the surface density of GCs in the vicinity to UCDs from what we would expect from the large scale distribution around NGC\,1399.

   \begin{figure}
   \centering
   \includegraphics[width=\hsize]{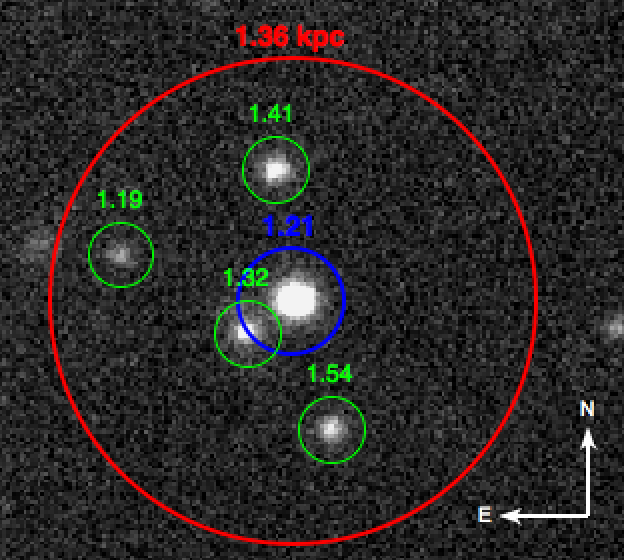}
      \caption{The blue circle marks the UCD with the original name Y4289 (\citealt{Richtler2008}), which is a confirmed member of Fornax. The UCD itself has an absolute magnitude of $M_V=-10.3$ and a color of $(V-I)=1.21\pm 0.05$. The four surrounding GC candidates \cite{Dirsch2003} are shown with green circles and they are labeled with their respective $C-T_{1}$ colors. A circle with a radius of 1.36\,kpc, centered on the UCD, is shown in red. This circle marks the estimated tidal radius from equation \ref{eq:tidal}. The cutout has been taken from the wide field VST Fornax survey imaging (Iodice et al. in prep.)}
         \label{fig:cand}
   \end{figure}
We also checked for significant GC clustering at large radii. For statistical studies the noise in the radial GC distribution (see fig. \ref{fig:gc1399}) can dilute an average signal. But for single UCDs highly significant clustering can still be detected. We found one UCD (blue circle in figure \ref{fig:cand}) with four GC candidates (green circles) within 1\,kpc radius (red circle). This UCD is located at RA= 03:38:30.768 and Dec =$-$35:39:56.88, 85\,kpc south of NGC1399. All 4 GC candidated lie well within the estimated tidal radius from equation \ref{eq:tidal} given as $r_{\rm tidal}$=1.36kpc, which is shown as red circle. The GC density around this UCD is $\Sigma=1.27$kpc$^{-2}$ although we only expect $\sim 0.04 \rm kpc^{-2}$ at these radial distances. Thus this is larger by a factor of $\sim 32$ than what would be expected from a simple radial distribution of GCs in the halo of NGC\,1399. In addition, all four GC candidates would belong to the blue (metal-poor) GC population if they are Fornax members, with an average of $C-T_{1}=1.37$. The UCD itself has a red color of $C-T_{1}=1.82$. This is in good agreement with our finding for the UCDs within 110\,kpc, for which the blue GCs are more clustered around UCDs than the red ones (see fig. \ref{fig:clust}). 
The position of this UCD in the outskirts of the halo of NGC\,1399 and its associated blue GC population make it a prime target for an infalling, nucleated dwarf galaxy in its early stages of disruption, where the original GC system might still be partially intact. However, it is very puzzling that no faint stellar envelope of tidal features have been found around this UCD. We will come back to this in the discussion.

\section{Discussion}

 \subsection{Surface brightness profiles of UCDs and their scaling relations}
 
One of the possible theories of UCD formation is that some of them are the remnant nuclei of larger dwarf ellipticals, being stripped of their stellar envelope through tidal interactions (e.g. \citealt{Bekki2003, Drinkwater2003}). In \cite{Pfeffer2013} the trajectories from simulations in the size-mass plane of such a process were shown. After being tidally stripped, they end up with sizes and magnitudes of UCDs, effectively having to cross the empty region in between the galaxy and the star cluster branch. Although it is noteworthy that these are idealized simulations which do not include a dark matter halo for the satellite galaxy, and thus apply only to galaxies that have already lost the majority of their dark matter. In section \ref{sec:34} we posed the question why do we not see more objects in between both branches being currently transformed if these objects are formed by tidal stripping? The predicted timescales of 2-3\,Gyrs for such a stripping process are not too short compared to a Hubble time to observe a certain fraction of these stripped galaxies during the process. This, of course, depends on the number of nucleated dEs available that are on the right disruptive trajectories within the galaxy cluster and on the timing of disruption events. Most probably, the majority of accretion events happened during the violent early phases of the galaxy cluster assembly.

In any case, for those UCDs that are the remnants of tidally stripped dwarf galaxies we expect that the overall size and luminosity of the disrupting dwarf galaxy will be decreased by the stripping of the faint stellar envelope that is left around the nuclear cluster. Most surface brightness distributions of the UCDs we found are well fitted with a Sérsic index between $2<n<6$. Most dEs obey an exponential $n=1$ surface brightness profile (when excluding the nuclear star cluster), whereas compact ellipticals and giant ellipticals usually require more centrally concentrated profiles with $n=4$. One explanation for this central concentration of the UCD profiles could be that 
 we see a superposition of the centrally concentrated nuclear star cluster (NSC) and the remains of the stellar envelope. While the stellar envelope is getting fainter and fainter during stripping, the luminosity from the NSC starts to dominate the profile more and more, and thus causes a more centrally concentrated light profile. 
  
Another possibility which could cause the appearance of faint envelopes around UCDs is the merging of star cluster complexes into one extended UCD. In \cite{Bruens2011} it has been shown with a numerical model that the products of such mergers at large galactocentric distances can have effective radii up to 80\,pc. In \cite{Fellhauer2005} the evolution of the surface brightness profiles of merging stellar superclusters are shown. Within the first 500\,Myrs such a star cluster complex can exhibit faint halo-like features up to a few hundred pc of distance. In their work they also followed the evolution of two super star cluster over 10 Gyrs. The initially relatively bright halos at 25-26mag/arcsec$^{-2}$ decrease quickly in size and surface brightness. 
After 10 Gyrs the faint envelope components have decreased to a surface brightness of 28-32 mag/arcsec$^{-2}$ at the maximum extension of $\sim$100\,pc. In addition, 
\cite{Fellhauer2005} showed that after evolving for 10\,Gyr these merged star cluster complexes are rather resembling the faint fuzzy star clusters than UCDs. Thus the origin of these faint envelopes from merged star clusters is rather unlikely, as none of our UCDs with envelopes show any signs of very young ages (<1\,Gyr) where a super star cluster envelope would still be bright enough to be detected. The predicted envelope magnitudes at 28-32mag/arcsec$^{-2}$ for a 10\,Gyr old star cluster complex are several orders of magnitude fainter than the $\mu$=26 \rm mag/arcsec$^{-2}$ envelopes we found. Of course, this does not exclude that several of the UCDs without detectable envelopes originated from merged star cluster complexes.

When we decomposed the UCD light profiles in a King core component and a Sérsic envelope we found that the Sérsic index of the envelopes takes values of $n<2$, which resembles more closely what is found for surface brightness profiles of dwarf ellipticals. In the size-magnitude diagram these remnant envelopes,
if regarded as isolated components, are located in the empty area between the galaxy and star cluster branch, which exactly is the transition area \cite{Pfeffer2013} predict for dEs that are being stripped.

An argument against UCDs being the nuclei of stripped dwarf galaxies is the location of the red extended UCDs on the color magnitude diagram. While blue extended UCDs agree well with the established nuclei locations in the CMD, the red UCDs are significantly fainter than what is expected from a galaxy nucleus.

The stripping scenario is also supported by the three works that found double profiles for individual UCDs  (\citealt{Evst2007, Chili2008, Strader2013}). Their core components resemble the scaling relations of nuclear star clusters, whereas the envelopes lie in the previously empty transition area between the galaxy and star cluster branch. But, as mentioned before, envelopes are also a transient feature of merged star cluster complexes.

\subsection{Tidal features}
We examined our sample of confirmed UCDs for direct observational evidence of ongoing tidal stripping as possible proof for the transformation of a dE galaxy into a UCD or the tidal disruption of a super star cluster. We found extended features around 11 UCDs with apparent sizes between 100 and 350\,pc. UCD-FORS\,2 shows the clearest evidence for a tidal feature, with two tails detected at $3\sigma$-level above the background, which have an apparent extent of 350\,pc. The theoretical tidal radius for this object is given as 444\,pc for its distance of 33.34\,pc to NGC\,1399, and is thus only slightly larger than the observed tidal features. For an object of this luminosity a tidal radius of 130\,pc would be reached at a pericenter distance of 10\,kpc. The 11 UCDs with extended features account for 11.3\% of the total population we analyzed in detail.

In \cite{Pfeffer2014} the number of stripped nuclei within $R<83$\,kpc and masses $M>2 \cdot 10^{6}M_{\odot}$ for the central parts of the Fornax cluster is predicted to be $n_{\rm nuclei}=11.6^{+5.7}_{-4.9}$. When we scale this value to the total surface of our three FORS fields we get a prediction of $n_{\rm nuclei}=19.07^{+9.4}_{-8.1}$ stripped nuclei in our observing fields. This is marginally consistent with our 11 UCDs with possibly direct evidence for tidal features. However, there are several factors to be considered when comparing those two numbers:
1) Some of our detections might be by-chance superpositions of low surface brightness background galaxies, although the calculated probability for that is very low.
2) Because tidal features are of low surface brightness, the number of UCDs with stripping evidence would rise if we were able to observe deeper. Our observations were able to detect tails down to a limiting surface brightness of $\mu=26$\,mag/arcsec$^{2}$. In \cite{Pfeffer2013} the simulated envelopes reach this magnitude level after being stripped for 700\,Myrs. Thus we miss stripping events that started more than $\sim$1\,Gyr ago. It is interesting to note that the surface brightness levels of the envelopes can act as sort of ``clock" for the disruption processes, giving hints on the age since the tidal disruption started. In this respect, our determined fraction of nuclei remnants puts a lower limit on the contribution of the tidal stripping as formation channel for UCDs.
3) Tidal tails only occur during the actual stripping period, which lasts of the order 2-3\, Gyr. In combination with our surface brightness limits this implies that the tidal tails are only observable during a very short time window of the stripping process. Since disruption events are distributed all over the galaxy cluster evolution time, and probably were more common during the main galaxy cluster assembly several Gyr ago, one expects to see only a small fraction of ongoing disruption events nowadays. \cite{Pfeffer2014} estimate that about 5\% of the available stripped nuclei candidates started stripping less than 2\,Gyr ago, which would mean that only one event should be seen within our FORS fields. Maybe UCD-FORS\,2 is the 'lucky' smoking-gun.

Taken all factors together we can only state that the low number of observed tidal features at the given surface brightness limit, $<$11 out of 97 UCDs, is not in contradiction with the expected number of stripped nuclei from simulations. In terms of contamination it is an upper limit for recently stripped dE,Ns, in terms of the total number of stripped nuclei it is a lower limit.

An alternative explanation for the appearance of the tail-like structures around UCD-FORS\,2 is that this object is a high mass globular cluster or a merged super star cluster that is  currently being disrupted. The Galactic GCs Palomar 14 (\citealt{Sollima2011}) and Palomar 5 (\citealt{Odenkirchen2001}) show tidal features of comparable size and symmetry to our object. It has to be noted, however, that both Palomar clusters have masses of M$\simeq 10^{4}M_{\odot}$, whereas our object is by a factor of 1000 more massive ($\sim10^{7}M_{\odot}$) at a comparable half-light radius. To strip stellar material from such a compact object requires a much stronger tidal field. The intermediate-age stellar populations detected in UCD-FORS2 (\citealt{Richtler2005}) might be hint that this UCD originated from a super star cluster complex that was formed a few Gyrs ago. On the other hand a youngish age would also be expected for a nuclear star clusters that was build up by recurrent star formation from infalling gas.

Based on our evidence we cannot clearly distinguish if this object is a massive globular cluster being stripped or the remnant of a nucleated galaxy. Although the high stellar density makes direct stripping very difficult, the Balmer lines and metallicity detected for this object in \cite{Richtler2005} point to the presence of a younger stellar population, which agrees well with a stripped GC complex. In any case, this is the first time that a significant tidal tail has been discovered around a UCD-like object of this magnitude.

\subsection{Spatial correlation of UCDs with globular clusters}  

In our statistical search for signatures of spatial clustering of GCs around UCDs within 10 to 110\,kpc from NGC\,1399, we found a local overabundance of GCs on scales of 0.5 to 1\,kpc around UCDs when compared to the global abundance of GCs in the halo. In total 48.5\% of our UCDs have an overdensity of globular clusters around them within 1\,kpc. Thus, it appears that UCDs can be divided into two distinct populations: 1) UCDs that harbour a population of close-by satellite points sources, most probably low mass star clusters;
2) UCDs that have the same statistical clustering properties as 'normal' globular clusters.

One possible explanation for the clustering signal of GCs around UCDs could be that UCDs formed as nuclei of a dwarf galaxy that harboured its own GC system. What we see is the remnant GC population of the ancestor galaxy before it was stripped to its nucleus.
The observed clustering is expected in a scenario in which globular clusters merge towards the center of the galaxy via dynamical friction in less than a Hubble time. In particular, the dynamical friction timescale is shortest for the most massive globular clusters of a dwarf galaxy. Thus in a migratory-merger model, high mass GCs would spiral into the nucleus first as they have dynamical friction timescales of $\tau<10$\,Gyrs.
Evidence of this process is provided in \cite{Lotz2001} who found a deficit of bright GCs in the central parts of dE galaxies. Their sample of dwarf galaxies between $-12>M_{V}>-16$ have their own GC population between 0-25 GCs per dwarf galaxy. This low number of observed GCs in a dwarf elliptical combined with the models of \cite{Arca-Sedda2014} leads to the prediction that a galaxy of $M=10^{9}M_{\odot}$ mass loses $~80\%$ of it's GC population over a Hubble time via dynamical friction, thus the observed number of GCs in old and potentially stripped objects should be fairly low, between 0-5 GCs. Even considering that a galaxy, which experienced strong mass loss after tidal stripping, has lost 95\% of its original mass, the remaining 5\% are still enough to harbour the NC, a companion GC and a faint remnant envelope.

In the literature, there are some examples of nucleated dwarf galaxies that possess only a few GCs (\citealt{Georgiev2009, Georgiev2010}). Recently, \cite{Karach2015} discovered a faint dwarf spheroidal, $M_{B}=-10.8$ that contains a central globular cluster with an extended faint stellar halo around it. Such objects might be the progenitors of the extended UCDs we find. The globular clusters around dwarf galaxies are in majority metal-poor and old, thus have a blue $(V-I)$ colour. Since the observed clustering signal is strongest for the blue GC subpopulation, this strengthens the view that some UCDs with blue companions are the stripped nuclei of metal-poor (dwarf) galaxies. This notion is also supported by object Y4289 for which we found 4 very blue (probably metal-poor) GCs within a radius of 1\,kpc and a red central UCD. The position of this UCD in the outskirts of the halo of NGC\,1399 and its associated blue GC population make it a prime target for an infalling, nucleated dwarf galaxy where the original GC system might still be partially intact. The large colour difference between the red nucleus and the blue GCs shows that the more metal-rich nucleus must have been metal-enriched through several SF episodes and most likely did not form similar to its surrounding metal-poor GCs. Alternatively, if this object is the result of a merger of a super star cluster complex, the blue colours of the surrounding GCs could be interpreted as youngish ages. But then one would expect that the central merged object (UCD) has the same colour, except it had been enriched in metals during the violent formation in the central part of the super star cluster complex. Nevertheless it is puzzling that assuming these are the remnant GCs of a disrupting dwarf galaxy, that we do not see a remnant stellar envelope. 
In our search for stellar envelopes we have a surface brightness limit of 26\,mag/arcsec$^{-2}$ and potential envelopes could be well below our detection limit, although bright dEs would have surface brightness of $\sim$ 23-25\,mag/arcsec$^{-2}$.

A very different explanation for GC clustering around UCDs could be the temporary capture and focussing of GCs by the potential well of a UCD that is moving through the halo of a large galaxy. The typical tidal radii for a $10^{6}M_{\odot}$ UCD in a galaxy like NGC\,1399 is given as 300\,pc, 700\,pc and 900\,pc for distances of 30, 50 and 70\,kpc to the center of the galaxy respectively. In this scenario, UCDs can keep objects bound for a considerable time if their pericenter passage is not too close to the central galaxy and if they move in a similar phase space as the surrounding GCs, which would be the situation for a super star cluster complex with others cluster formation around it. In the outskirts of a halo like NGC1399 the tidal radius of an UCD can grow up to 1\,kpc or larger and is in good agreement with the radial scale on which we find the clustering signal.

An alternative explanation for the clustering of GCs around UCDs could be that those UCDs initially formed in a star cluster complex that then subsequently merged to create a UCD (see e.g. \citealt{Fellhauer2002,Bruens2009, Bruens2011}). Massive star cluster complexes are common in gas-rich mergers (e.g. The Antennae, \citealt{Whitmore1999}). The overdensites of star clusters around the merged super star cluster then are the remnants of this process. In the simulations of \cite{Bruens2011, Bruens2009} they predict that these kind of star cluster complexes would merge within 1\,Gyr, although in some cases substructure can last up to 5\,Gyr in their simulations. Thus, in order to still detect a remaining GC of this complex, which has not merged yet, the UCD and its surrounding GCs would have to have rather young ages and probably a high metallicity. Currently, spectroscopic observations of UCDs and GCs in Fornax indicate that most of them have an age older than 8\,Gyr (\citealt{Mieske2006}). Some  Fornax UCDs (\citealt{Chili2011}, \citealt{Chili2008}, \citealt{Richtler2005}), however, show intermediate ages, which could resemble a recently merged super star cluster complex. In contrast to the star cluster scenario, the majority of the companion GCs around UCDs are most probably metal-poor (see above) which points to a dwarf galaxy origin. Thus, we consider both the cluster complex merger scenario and the tidally stripped nucleated galaxy scenario as viable formation channels for Fornax UCDs with companion objects, and to discriminate between them one has to carefully compare the metallicities and ages of the nuclei and the companions. 
 
In conclusion, we interpret the structural parameters of the studied UCDs and their faint envelopes as well as the discovered tidal features as indication that the stripping of a dwarf galaxy is a viable formation channel of at least a fraction of UCDs. Homogeneous and deep observations targeted on finding and confirming the tidal features and improving the sample of UCDs with detailed structural parameters are necessary to draw definite conclusions about their origins and the contribution of each formation channel to the number counts of UCDs.

For the clustering of GCs around UCDs, it is important to confirm our finding with larger datasets in other galaxy clusters.
The association of UCDs and their companion GCs can be probed by radial velocity and stellar population measurements. This is one of the key observations to distinguish whether these objects are in fact the remnants of disrupted galaxies or the remnants of merging star clusters.

\begin{acknowledgements}
Firstly we would like to thank the referee Igor Chilingarian for his useful comments that improved the paper. We also wish to thank Thorsten Lisker and Carolin Wittmann for their helpful comments on the manuscript and Pavel Kroupa for fruitful discussions. We thank Andi Burkert for discussing the possible explanations for the GC clustering. We also want to thank Enrica Iodice for kindly providing us the VST image around UCD Y4289 and thank Matthieu Bethermin for his useful advice on the statistics. TR acknowledges financial support from the BASAL Centro de Astrofisica y Tecnologias Afines (CATA) PFB-06/2007. He also thanks ESO/Garching for hospitality.
 \end{acknowledgements} 

\bibliographystyle{aa}
\bibliography{fnx_ucd_arxiv}

\newpage
\begin{appendix}

\section{Tables of UCD properties}

\onecolumn

\begin{landscape}
 \begin{table*}
\caption{UCD and companion properties of all 19 objects for which have a faint point source was found within r<300pc (see figure \ref{fig:comp}). In column 1 the reference used in figure \ref{fig:comp} is shown. In column 2 and 3 the name being used in this paper and the original name are shown respectively. Column 4 and 5 give their positions in R.A. and Dec. In column 6 and 7 the V magnitude and colour of the UCD are shown. In column 8 the magnitude if the possible companion is given. In column 9 the distance of the companion to the UCD is given in pc. The distance of the UCD to the center of NCG1399 is given in column 10. In column 11 we show the estimate of the tidal radius of this UCD for it's specific position using the formula given in \ref{eq:tidal}. The second to last column gives the fraction between $dist_{comp}/r_{tidal}$. Thus if this fraction is smaller than 1.0 the companion lies within the tidal radius of the UCD host object. Which is the case for 16 our of our 19 UCDs.}
\begin{tabulary}{1.0\textwidth}{cccccccccccccc}
\hline \hline
Ref.  & Name  & Name$_{\rm alt}$ & R.A. & DEC. & $V_{UCD}$ & $(V-I)$ & $V_{comp.}$ & $dist_{comp}$ & $dist_{NGC1399}$ & $r_{tidal}$ & $dis_{comp}/r_{tidal}$ & $v$  \\
& & & (h:m:s) & ($^{\circ}$:':'') & (mag) & (mag) & (mag) & (pc) & (kpc) & pc & & (km s$^{-1}$) \\
\hline 
a) & UCD-FORS 1 &  1\_0630 & 3:38:56.14 & -35:24:49.0 &  20.35 &  1.02 & 25.21 &  108 & 32.86   &    590 & 0.18 & 666  $\pm$ 48 \\
b) & UCD-FORS 13 &  Y446  & 3:38:33.82 & -35:25:57.0 & 21.01 & 1.03  & 25.01   &    211 & 8.07   &    119 & 1.77 & 1224 $\pm$   221 \\
c) & UCD-FORS 13 &  Y446  & 3:38:33.82 & -35:25:57.0 & 21.01 & 1.03  & 23.80 &  307 & 8.07  &   119  & 2.58 &1224 $\pm$   221 \\
d) & UCD-FORS 13 &  Y446  & 3:38:33.82 & -35:25:57.0 & 21.01 & 1.03  & 24.59   &  241 & 8.07  &    119 & 2.03 & 1224 $\pm$   221 \\
e) & UCD-FORS 20 &  77.002  & 3:38:38.11 & -35:26:46.7 & 21.02 & 1.03 & 24.81 &  119 & 10.32 & 151 & 0.79 & 1788  $\pm$ 58  \\
f) & UCD-FORS 32 &  UCD6  & 3:38:05.04 & -35:24:09.7 &  18.93 & 1.13 &  23.20  & 263 &  31.30    &   985 & 0.27 & 1220   $\pm$  45 \\
g) & UCD-FORS  33 &  NTT414  & 3:38:09.72 & -35:23:01.2 & 19.61 & 0.99 &  23.78 &  206  &   31.03   &  686 & 0.30 & 1607 $\pm$  141 \\
h) & UCD-FORS  35 &  UCD32  & 3:38:16.70 & -35:20:23.1 & 20.09 & 0.99 &  23.64  &  278  & 39.24     &   748 & 0.37 & 1439 $\pm$ 23 \\
i) & UCD-FORS  36 &  UCDm & 3:38:06.48 & -35:23:03.8 & 20.01 & 1.0 &  25.12   &  283 &  33.53    &   659 & 0.43 & 1442 $\pm$  123 \\
j) & UCD-FORS  37 &  UCD36 & 3:38:23.23 & -35:20:00.6  & 20.22 & 1.12  &   24.62 & 139  & 39.35 &   820 & 0.17 & 1347 $\pm$ 61 \\
k) & UCD-FORS  52 &  2\_2127 & 3:38:11.69 & -35:27:16.2 & 20.93 & 1.16 & 23.77   &  154 &  19.53   &   349 & 0.44 & 1443  $\pm$ 131\\
l) & UCD-FORS  57 &  Y3905 & 3:38:23.28 & -35:26:32.7 & 21.27 & 1.09  &  24.62 & 224 &  7.01   & 101 & 2.22 &  1504 $\pm$  12 \\
m) & UCD-FORS 64 &  K1042m & 3:38:36.88 & -35:25:43.8 & 21.37 & 1.23  &  22.71  & 151 &  11.43  & 193 & 0.78 & 1328  $\pm$  90 \\
n) & UCD-FORS 69 &  Y9410 & 3:38:27.35 & -35:25:37.5 & 21.30 & 1.16  &  25.08 &  174 & 8.02  &  127 & 1.37 & 1392 $\pm$ 32 \\
o) & UCD-FORS  70 &  Y4081 & 3:38:26.63 & -35:25:34.0 & 21.39 & 1.22 &  25.50  &  100  & 8.56 &  143 & 0.70 & 1442 $\pm$  22 \\
p) & UCD-FORS  71 &  Y10048 & 3:38:35.23 & -35:25:39.2 & 21.57 & 1.09  &  25.11  &   63  & 10.34 &  136 & 0.46 & 1602 $\pm$ 30 \\
q) & UCD-FORS  76 &  80.039 & 3:38:21.34 & -35:24:35.6 & 21.63 & 1.13 &  23.93   & 211 &  16.03  &   220  & 0.96 &  900      $\pm$ 66\\
r) & UCD-FORS  80 &  K1040 & 3:38:28.84 & -35:25:00.7 & 21.93 & 1.17  &  24.92  &  124  &  11.18     &  149 & 0.83 &  1481  $\pm$     57 \\
s) & UCD-FORS  81 &  UCD2 & 3:38:06.29 & -35:28:58.8 & 19.31 & 1.13 &   24.47  &  193 &  27.71   &  776 & 0.24 & 1249 $\pm$  37 \\
t) & UCD-FORS 92 &  Y7623 & 3:38:16.65 & -35:29:35.2 & 21.59 & 1.05  &  24.22 &  177 & 19.77 &   248  & 0.71 & 1676 $\pm$  50 \\
u) & UCD-FORS  96 &  89.038 &  3:38:11.93 & -35:32:01.7 & 21.45 & 0.97  &   23.72  &  291  &  33.58 & 417 & 0.70  &  1350 $\pm$ 78\\
\end{tabulary}
\label{companiondat}
\end{table*}
\end{landscape}
 
 \onecolumn
 
\begin{longtable}{cccccc}
\caption{Here those UCDs from the original sample of 97 in the FORS fields are shown, which are not yet lister in table \ref{ucddat} or as a companion hosting UCD in table \ref{companiondat}. Column one gives their running name in our notation, column to their alternative name from the literature, which follow the same convention as explained in table \ref{ucddat}. Column 3 and 4 give the R.A. and Dec. of the UCDs. In column 5 and 6 the V-magnitudes as well as V-I colors are shown.}\\
\hline 
\hline
 Name  & Name$_{\rm alt}$ & R.A. & DEC. & $V_{UCD}$ & $(V-I)$ \\
& & (h:m:s) & ($^{\circ}$:':'') & (mag) & (mag) \\
\hline
\endfirsthead
\caption{continued.}\\
\hline \hline
 Name  & Name$_{\rm alt}$ & R.A. & DEC. & $V_{UCD}$ & $(V-I)$ \\
& & (h:m:s) & ($^{\circ}$:':'') & (mag) & (mag) \\
\hline
\endhead
\hline 
\endfoot
UCD-FORS 3 & 1\_2103 & 3:38:57.38  & -35:24:50.8  &     20.82 $\pm$  0.11  &   1.14   $\pm$    0.07 \\
UCD-FORS 5 & 1\_064 & 3:38:49.78 &  -35:23:35.5  &   20.99  $\pm$   0.08   &       1.18   $\pm$   0.05 \\
UCD-FORS 6 & 76.059 & 3:38:46.90 & -35:23:48.8    &   21.10   $\pm$  0.08 &     1.01   $\pm$   0.08 \\
UCD-FORS 7 & 75.085 & 3:38:50.40 & -35:22:07.7    &   21.46  $\pm$   0.08   &     1.23   $\pm$  0.08 \\
UCD-FORS 8 & Y5100 & 3:38:48.93 & -35:21:22.6     &  21.73 $\pm$      0.10   &     1.36   $\pm$  0.05 \\
UCD-FORS 9 & 78.070 & 3:38:42.48 & -35:26:12.5    &   21.88 $\pm$  0.09   &    1.20   $\pm$   0.05 \\
UCD-FORS 10 & UCD41 &  3:38:29.04   & -35:22:56.6   &    19.98  $\pm$   0.07   &      1.16   $\pm$   0.04 \\
UCD-FORS 11 & 0\_2030 & 3:38:28.34   & -35:25:38.3   &    20.08 $\pm$     0.11   &     1.10   $\pm$   0.05 \\
UCD-FORS 12 & Y446 & 3:38:30.72   & -35:24:40.7    &   21.01 $\pm$      0.10   &     1.03   $\pm$   0.06 \\
UCD-FORS 14 & 1\_058 & 3:38:39.31  &  -35:27:06.5   &    20.72 $\pm$      0.15   &        1.08   $\pm$    0.13 \\
UCD-FORS 15 & AAT38 & 3:38:37.97   & -35:23:33.0    &   20.90 $\pm$      0.10   &       1.19   $\pm$    0.13 \\
UCD-FORS 16 & 0\_2074 & 3:38:35.66   & -35:27:15.5   &    21.03 $\pm$     0.12   &      1.11   $\pm$   0.07 \\
UCD-FORS 17 & 76.080 & 3:38:40.99   & -35:22:41.9    &   21.27 $\pm$     0.08   &       1.15   $\pm$   0.07 \\
UCD-FORS 18 & Y4735 & 3:38:40.24  &  -35:21:33.2    &   21.48 $\pm$      0.15   &      1.15   $\pm$   0.07 \\
UCD-FORS 19 & K1044a & 3:38:40.20   & -35:27:00.7  &     21.30 $\pm$    0.08   &      0.89   $\pm$   0.05 \\
UCD-FORS 21 & Y4654 & 3:38:38.75   & -35:25:42.9    &   21.64 $\pm$     0.04   &      1.17   $\pm$  0.06 \\
UCD-FORS 22 & UCD27 & 3:38:10.34   & -35:24:06.1    &   19.70 $\pm$     0.11   &    1.10   $\pm$   0.06 \\
UCD-FORS 23 & Y99071 & 3:38:08.64   & -35:23:51.8   &    22.02   $\pm$   0.10   &       4.89   $\pm$  0.04 \\
UCD-FORS 24 & 80.056 & 3:38:19.73   & -35:23:40.6   &    20.66 $\pm$    0.09   &     0.97   $\pm$   0.08 \\
UCD-FORS 25 & 0\_2032 & 3:38:30.22  &  -35:21:31.0   &    20.86 $\pm$    0.08   &       1.04   $\pm$   0.08 \\
UCD-FORS 26 & 81.049 & 3:38:08.35   & -35:23:56.0   &    22.12 $\pm$     0.53   &      1.02   $\pm$   0.24 \\
UCD-FORS 27 & 81.041 & 3:38:07.06  &  -35:24:28.8    &   21.60 $\pm$   0.08   &      1.26   $\pm$   0.06 \\
UCD-FORS 28 & 81.098 & 3:38:07.66  &  -35:20:51.4   &    21.93 $\pm$   0.07   &     0.94   $\pm$   0.05 \\
UCD-FORS 29 & 81.066 & 3:38:06.31  &  -35:22:48.4   &    21.85 $\pm$   0.08   &      0.94   $\pm$   0.05 \\
UCD-FORS 30 & 0\_2030 & 3:38:28.34  &  -35:25:38.3   &    20.08 $\pm$    0.11   &      1.10   $\pm$  0.05 \\
UCD-FORS 31 & NTT407 & 3:38:04.17  &  -35:25:26.6   &    20.24 $\pm$   0.17   &      0.91   $\pm$   0.11 \\
UCD-FORS 34 & Y446 & 3:38:30.72  &  -35:24:40.7    &    21.01   $\pm$   0.10   &      1.03   $\pm$  0.06 \\
UCD-FORS 38 & 0\_2063 & 3:38:19.08  &  -35:26:37.3   &    20.87 $\pm$     0.05   &       1.06   $\pm$   0.05 \\
UCD-FORS 39 & 0\_2027 & 3:38:19.49  &  -35:25:52.3    &   21.08   $\pm$  0.09   &      1.25   $\pm$   0.06 \\
UCD-FORS 40 & 0\_2089 & 3:38:17.09  &  -35:26:30.8   &   21.00 $\pm$      0.13   &      1.13   $\pm$   0.04 \\
UCD-FORS 41 & K1026 & 3:38:14.25  &  -35:26:43.9   &    20.91  $\pm$  0.08   &      1.11   $\pm$   0.06 \\
UCD-FORS 42 & 80.028 & 3:38:26.47  &  -35:25:21.0   &    21.31 $\pm$     0.09   &      1.08   $\pm$   0.06 \\
UCD-FORS 43 & Y7797 & 3:38:26.40  &  -35:24:25.6    &   21.31 $\pm$  0.06   &      1.08   $\pm$  0.05 \\
UCD-FORS 44 & Y4222 & 3:38:29.52  &  -35:25:08.5    &   21.31 $\pm$  0.10   &     1.21   $\pm$  0.06 \\
UCD-FORS 46 & Y9320 & 3:38:19.99  &  -35:26:44.0    &   21.29 $\pm$    0.10   &     1.08   $\pm$   0.06 \\
UCD-FORS 47 & 80.027 & 3:38:26.28  &  -35:25:25.0  &     21.38 $\pm$   0.08   &     1.24   $\pm$   0.05 \\
UCD-FORS 49 & 80.035 & 3:38:18.22  &  -35:24:54.0   &    21.30 $\pm$    0.09   &      1.08   $\pm$   0.06 \\
UCD-FORS 51 & 82.040 & 3:38:10.32  &  -35:26:31.9   &    21.86 $\pm$     0.38   &     1.22   $\pm$   0.19 \\
UCD-FORS 53 & Y4507 & 3:38:35.48  &  -35:25:29.6    &   21.41 $\pm$   0.04   &    1.18   $\pm$   0.05 \\
UCD-FORS 54 & 82.029 & 3:38:05.66  &  -35:26:46.7  &     22.04 $\pm$   0.41   &      1.22   $\pm$    0.09 \\
UCD-FORS 55 & Y3866 & 3:38:22.41  &  -35:26:33.2    &   21.15 $\pm$   0.11   &    1.10   $\pm$    0.07 \\
UCD-FORS 58 & Y7698 & 3:38:20.60  &  -35:26:11.3   &    21.62 $\pm$   0.06   &       1.15   $\pm$   0.05 \\
UCD-FORS 59 & Y7967 & 3:38:35.78  &  -35:25:34.1   &    21.62 $\pm$  0.02   &      1.01   $\pm$  0.04 \\
UCD-FORS 60 & Y7935 & 3:38:33.86  &  -35:25:21.9   &    21.72 $\pm$   0.25   &     0.83   $\pm$   0.05 \\
UCD-FORS 61 & Y9354 & 3:38:23.21  &  -35:25:29.7   &    21.94 $\pm$    0.07   &       1.32   $\pm$    0.13 \\
UCD-FORS 62 & K1031 & 3:38:21.54   & -35:26:16.1   &    21.89 $\pm$   0.08   &    1.01   $\pm$  0.04 \\
UCD-FORS 63 & K1032a & 3:38:21.69  &  -35:25:14.9  &     21.92 $\pm$    0.08   &      1.20   $\pm$   0.04 \\
UCD-FORS 65 & 80.045 & 3:38:22.85  &  -35:24:23.0    &   22.18 $\pm$      0.13   &    0.78   $\pm$   0.27 \\
UCD-FORS 66 & UCD33 & 3:38:17.47  &  -35:33:04.0   &    20.40 $\pm$   0.06   &     0.99   $\pm$   0.05 \\ 
UCD-FORS 67 & 0\_2023 & 3:38:12.70  &  -35:28:57.0   &    20.85 $\pm$    0.08   &         1.10   $\pm$   0.08 \\
UCD-FORS 68 & 0\_2026 & 3:38:18.89  &  -35:32:23.3   &    20.98 $\pm$   0.09   &     1.11   $\pm$    0.07 \\
UCD-FORS 72 & 2\_089 & 3:38:14.02  &  -35:29:43.1   &    20.96 $\pm$   0.08   &     1.04   $\pm$    0.06 \\
UCD-FORS 73 & Y3786 & 3:38:20.81  &  -35:34:26.9   &    21.06 $\pm$   0.03   &      0.96   $\pm$  0.04 \\
UCD-FORS 74 & 2\_2100 & 3:38:00.17  &  -35:30:08.3   &    21.12 $\pm$    0.09   &      1.05   $\pm$    0.07 \\
UCD-FORS 75 & 91.113 & 3:38:08.16  &  -35:27:52.2   &    21.08 $\pm$   0.09   &    1.01   $\pm$   0.05 \\
UCD-FORS 77 & 89.055 & 3:38:22.44  &  -35:30:49.0   &    21.14 $\pm$    0.10   &     0.96   $\pm$    0.06 \\
UCD-FORS 78 & Y7523 & 3:38:08.80  &  -35:32:25.5    &   21.53 $\pm$   0.03   &     1.80    $\pm$  0.03 \\
UCD-FORS 79 & 90.074 & 3:38:08.78  &  -35:29:39.5    &   21.17 $\pm$     0.07   &     0.95   $\pm$  0.05 \\
UCD-FORS 82 & 92.060 & 3:38:04.68  &  -35:30:07.9  &     21.17 $\pm$     0.08   &      1.16   $\pm$   0.05 \\
UCD-FORS 83 & AAT21 & 3:38:13.07  &  -35:31:07.4   &    21.11 $\pm$      0.11   &    1.09   $\pm$   0.12 \\
UCD-FORS 85 & 90.077 & 3:38:11.30  &  -35:29:31.2   &    21.43 $\pm$     0.09   &    1.08   $\pm$   0.05 \\
UCD-FORS 86 & 90.015 & 3:38:14.66  &  -35:33:25.6   &    21.31 $\pm$    0.08   &      1.22   $\pm$   0.05 \\
UCD-FORS 87 & gc156 &  3:38:13.56  &  -35:28:56.3   &    21.28 $\pm$     0.02   &     1.15   $\pm$   0.03 \\
UCD-FORS 88 & 91.083 & 3:37:58.75  &  -35:29:32.3   &    21.43 $\pm$     0.08   &     1.04   $\pm$   0.06 \\
UCD-FORS 89 & 91.109 & 3:38:04.44  &  -35:28:11.6    &   21.42 $\pm$    0.07   &    1.19   $\pm$   0.05 \\
UCD-FORS 90 & 89.037 & 3:38:15.46  &  -35:32:04.2   &    21.43 $\pm$     0.08   &      1.04   $\pm$   0.05 \\
UCD-FORS 91 & Y7703 & 3:38:20.99  &  -35:30:13.1    &   21.55 $\pm$      0.11   &     1.03   $\pm$  0.05 \\
UCD-FORS 93 & Y3603 & 3:38:16.49  &  -35:31:07.7    &   21.59 $\pm$      0.10   &      1.01   $\pm$  0.04 \\
UCD-FORS 94 & 90.044 & 3:38:18.31  &  -35:31:34.7   &    22.20 $\pm$      0.18   &    0.94   $\pm$   0.04 \\
UCD-FORS 95 & 89.042 & 3:38:13.25  &  -35:31:43.0   &    21.85 $\pm$     0.08   &     0.86   $\pm$  0.13 \\
UCD-FORS 97 & K1022 & 3:38:09.21  &  -35:35:07.3    &   20.85 $\pm$     0.07   &       1.31   $\pm$   0.05 \\
\label{alldat}
\end{longtable}

\end{appendix}

\end{document}